\documentclass[12pt,preprint]{aastex}
\usepackage{graphicx}

\slugcomment{Accepted for publication in ApJ.}

\begin{document}

\title{Solutions for 10,000 Eclipsing Binaries in the Bulge Fields of OGLE~II Using DEBiL}

\author{Jonathan Devor}

\affil{Harvard-Smithsonian Center for Astrophysics, 60 Garden Street, Cambridge, MA 02138}

\email{jdevor@cfa.harvard.edu}

\begin{abstract}
We have developed a fully-automated pipeline for systematically
identifying and analyzing eclipsing binaries within large datasets
of light curves. The pipeline is made up of multiple tiers which
subject the light curves to increasing levels of scrutiny. After
each tier, light curves that did not conform to a given criteria
were filtered out of the pipeline, reducing the load on the
following, more computationally intensive tiers. As a central
component of the pipeline, we created the fully automated Detached
Eclipsing Binary Light curve fitter (DEBiL), which rapidly fits
large numbers of light curves to a simple model. Using the results
of DEBiL, light curves of interest can be flagged for follow-up
analysis. As a test case, we analyzed the 218699 light curves
within the bulge fields of the OGLE~II survey and produced 10862
model fits\footnote{The list of OGLE~II bulge fields model
solutions as well as the latest version of the DEBiL source code
are available online at: \texttt{http://cfa-www.harvard.edu/$\sim
$jdevor/DEBiL.html}}. We point out a small number of extreme
examples as well as unexpected structure found in several of the
population distributions. We expect this approach to become
increasingly important as light curve datasets continue growing in
both size and number.
\end{abstract}


\section{Introduction}

Light curves of eclipsing binary star systems provide the only
known direct method for measuring the radii of stars without
having to resolve their stellar disk. These measurements are
needed for better constraining stellar models. This is especially
important for such cases as low-mass dwarfs, giants, pre-main
sequence stars and stars with non-solar compositions, for which we
currently have a remarkably small number of well-studied examples.
Other important benefits of locating binary systems include more
accurate calibration of the local distance ladder
\citep{Paczynski96,Kaluzny98}, constraining the low mass IMF, and
discovering new extrasolar planets\footnote{Although there are
additional complications, extrasolar planets can be seen as the
limiting case where one of the binary components has zero
brightness.}. In order to obtain measurements of the stars' radii
and masses, one needs to incorporate both the light curve
(photometric observations) and radial velocities (spectroscopic
observations) of the system. Since making large-scale
spectroscopic surveys is significantly more difficult than making
photometric surveys, it is far more efficient to begin with a
photometric survey and later follow-up with spectroscopic
observation only on systems of interest.

During the past decade, there have been numerous light curve
surveys (e.g. OGLE: \citet{Udalski94}; EROS: \citet{Beaulieu95};
DUO: \citet{Alard97}; MACHO: \citet{Alcock98}) that take advantage
of advances in photometric analysis, such as difference image
analysis \citep{Crotts92,Phillips95,Alard98}. The original goal of
many of these surveys was not to search for eclipsing binaries,
but rather to search for gravitational microlensing events
\citep{Paczynski86}. Fortunately, the data derived from these
surveys are ideal for eclipsing binary searches as well. More
recently, there have also been mounting efforts to create
automated light curve surveys (e.g. ROTSE: \citet{Akerlof00}; HAT:
\citet{Bakos04} ; TrES: \citet{Alonso04}) using small robotic
telescopes for extrasolar planet searches. The upcoming large
synoptic surveys (e.g. Pan-STARRS: \citet{Kaiser02}; LSST:
\citet{Tyson02}), spurred by the decadal survey of the National
Academy of Sciences, are expected to dwarf all the surveys that
precede them. Put together, these surveys provide an exponentially
growing quantity of photometric data \citep{Szalay01}, with a
growing fraction becoming publicly available.

\section{Motivation}
\label{secMotivation}

For over 30 years many codes have been developed with the aim of
fitting increasingly complex models to eclipsing binary light
curves (e.g. \citet{ Wilson71,Nelson72,Wilson79,Etzel81}). These
codes have had great success at accurately modeling the observed
data, but also require a substantial learning curve to fully
master their operation. The result is that up till now there have
been no comprehensive catalogue of reliable elements for eclipsing
binaries \citep{Cox00}.

\begin{figure}
\plotone{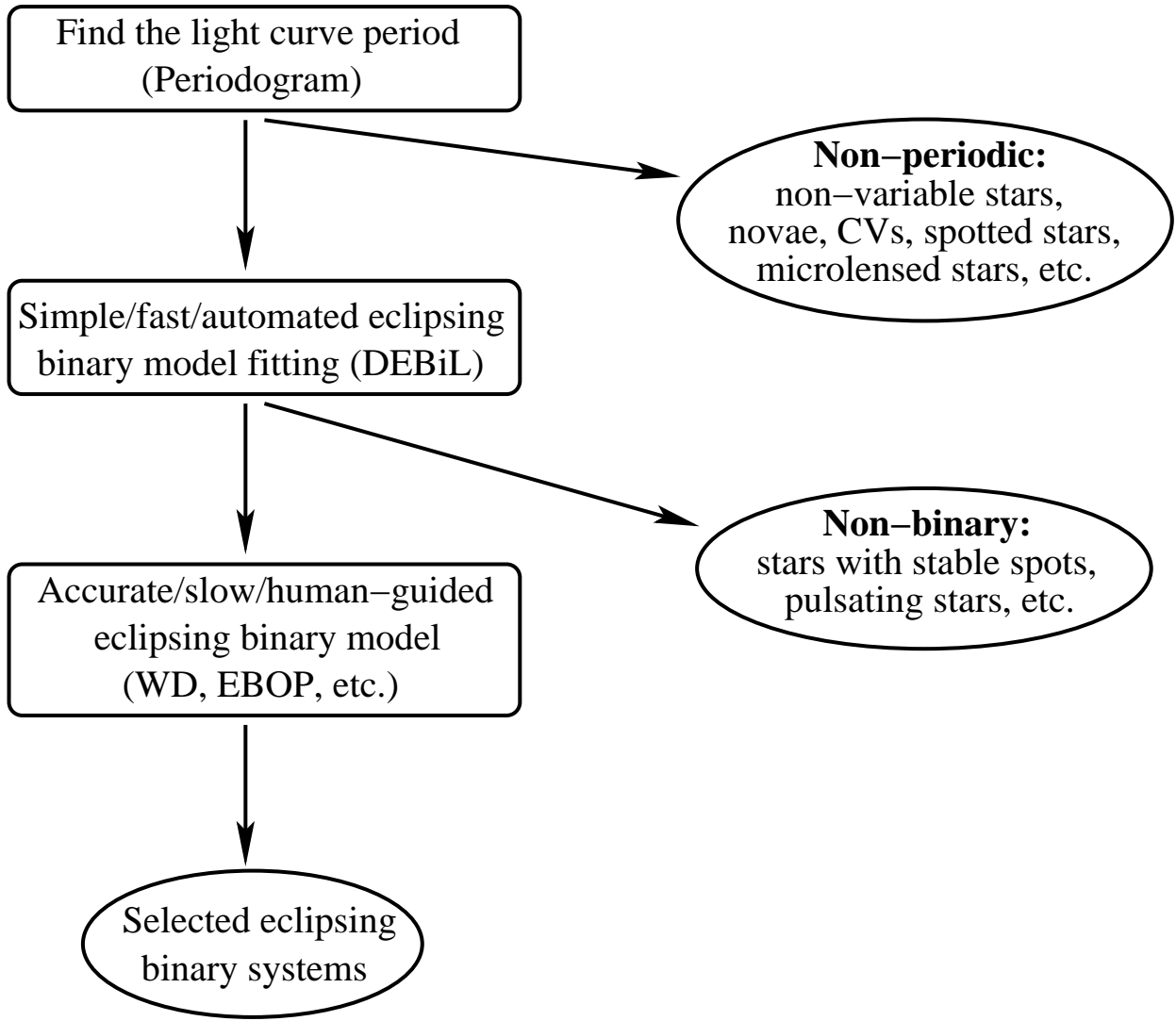}
\caption{Diagram of a multi-tiered model-fitting pipeline.}
\label{figBoxChart}
\end{figure}

In order to take full advantage of the large-scale survey
datasets, one must change the traditional approach of manual light
curve analysis. The traditional method of painstakingly fitting
models one by one is inherently limited by the requirement of
human guidance. Ideally, fitting programs should be both
physically accurate and fully automated. Many have cautioned
against full automation \citep{Popper81a,Etzel91,Wilson94} since
it is surprisingly difficult, without the aid of a human eye, to
recognize when a fit is ``good''. In addition, it is essentially
impossible to resolve certain parameter degeneracies without
a~priori knowledge and extensive user experience. Despite these
challenges, there have been a small number of pioneering attempts
at such automated programs \citep{Wyithe01, Wyithe02}. However,
the large numerical requirements of these programs make it
computationally expensive to perform full fits (i.e. without
having some parameters set to a constant) of large-scale datasets.
Moore's law, which stated in effect that CPU speed doubles every
18 months, cannot in itself solve this problem, since the quantity
of data to be analyzed is also growing at a similar exponential
rate \citep{Szalay01}. Instead, we advocate replacing the approach
of using monolithic automated fitting programs with a multi-tiered
pipeline (fig.~\ref{figBoxChart}). In such a pipeline, a given
light curve is piped through a set of programs that analyze it
with increasing scrutiny at each tier. Light curves that are
poorly fit or don't comply with set criteria are filtered out of
the pipeline, thus passing a far smaller number of candidates on
to the following, more computationally demanding tier. Such a
pipeline, coupled with efficient analysis programs, can increase
the effective speed of fitting models by a few orders of magnitude
compared to monolithic fitting programs. This approach makes it
practical to perform full fits of the largest light curve
datasets, with only moderate computational resources. Using a
single-CPU SUN UltraSPARC 5 workstation (333 MHz), the average
processing time of our pipeline was $\sim$1 minute per light
curve. Where each light curve typically contains a few hundred
photometric observations\footnote{The processing time scales
linearly with the average number of observations in the light
curves.}. We must emphasize that even at these speeds, we still
need a few CPU-months to fully process an OGLE-like survey
containing $10^5$ light curves. In about a decade, the large
synoptic surveys are slated to create datasets that are more than
4 orders of magnitude larger than that \citep{Tyson02}.

\placefigure{figBoxChart}

The Detached Eclipsing Binary Light curve fitter (DEBiL) is a
program we created to serve as an intermediate tier for such a
multi-tiered pipeline. DEBiL is a fully automated program for
fitting eclipsing binary light curves, designed to rapidly fit a
large dataset of light curves in an effort to locate a small
subset that match given criteria. The matched light curves can
then be more carefully analyzed using traditional fitters.
Conversely, one can use DEBiL to filter out the eclipsing binary
systems, in order to study other periodic systems (e.g. spotted or
pulsating stars). In order to achieve speed and reliable
automation, DEBiL employs a simple model of a perfectly detached
binary system: limb darkened spherical stars with no reflections
or third light, in a classical 2-body orbit. Given the system's
period and quadratic limb darkening coefficients (the default is
solar limb darkening), the DEBiL fitter will fit the following
eight parameters:

\begin{quote}
\begin{itemize}
\item Radius of primary star
\item Radius of secondary star
\item Brightness of primary star
\item Brightness of secondary star
\item Orbital eccentricity
\item Orbital inclination
\item Epoch of periastron
\item Argument of periastron
\end{itemize}
\end{quote}

Since we don't have an absolute length-scale, we measure both
stars' radii as a fraction of the sum of their orbital semimajor
axes (a).

This simple model allows the use of a nimble convergence
algorithm that is comprised of many thousands of small steps that
scour a large portion of the problem's phase space. Admittedly,
this model can only give a crude approximation for semidetached
and contact binaries, but it can easily identify such cases and
can flag them for an external fitting procedure. In addition to
pipeline filtration, DEBiL can also provide an initial starting
guess for these external fitters, a task that usually has to be
performed manually.

\section{Method}

This section describes our implementation of a multi-tiered light
curve-fitting pipeline. We fine-tuned and tested our design using
light curves from the bulge fields of the second phase Optical
Gravitational Lensing Experiment (OGLE~II)
\citep{Udalski97,Wozniak02}. Our resulting pipeline consists of
the following six steps:

\begin{quote}
First tier: Periodogram
\begin{enumerate}
\item [(1)] Find the light curve period
\item [(2)] Filter out non-periodic light curves
\end{enumerate}

Second tier: DEBiL
\begin{enumerate}
\item [(3)] Find an ``initial guess'' for the eclipsing binary
model parameters
\item [(4)] Filter out non-eclipsing systems (i.e. pulsating stars)
\item [(5)] Numerically fit the parameters of a detached
eclipsing binary model
\item [(6)] Filter out unsuccessful fits
\end{enumerate}
\end{quote}

For OGLE II data, we found that about half of the total CPU-time
was spent on step (1) and about half on step (5). The remaining
steps required an insignificant amount of CPU-time. Note that step
(5), which typically takes more than 10 times longer to run than
step (1), was only run on less than $10\%$ of the light curves. By
filtering out more than $90\%$ of the light curves at earlier
steps, the pipeline was able to run $\sim$10 times faster than it
would have been able to otherwise. Both steps (1) and (5) can
themselves be speeded up, but at a price of lowering their
reliability and accuracy. The third tier, in which the light
curves of interest are fitted using physically accurate models, is
dependent on the research question being pursued and won't be
further discussed here.

\subsection{The First Tier -- Finding the Period}

Step (1) is performed using an ``off-the-shelf'' period search
technique. All the periodogram algorithms that we have tested give
comparable results, and we adopted an analysis of variance
\citep{SchwarzenbergCzerny89, SchwarzenbergCzerny96} as it appears
to do a good job handling the aliasing in OGLE light curves. In
our implementation, we scanned periods from 0.1 days up until the
full duration of the light curve ($\sim$1000 days for OGLE~II). We
then selected the period that minimizes the variance around a
second order polynomial fit within 8 phase bins. Aliases pose a
serious problem for period searches since they can prevent the
detection of weak periodicities, or periods that are close to an
alias. The ``raw'' period distribution of the OGLE~II light curves
showed aliases with a typical widths ranging from 0.001 days for
the shortest periods, up to 0.04 days for longest periods. We
suppressed the 12 strongest aliases, over which the results were
dominated by false positives, and had the period finder return the
next best period. Fewer than $1\%$ of the true light curve periods
are expected to have been affected by this alias suppression.
Finally, once a period was located, rational multiples of it, with
numerators and denominators of 1 through 19 were also tested to
see if they provide better periods.

Step (2) filters out all the light curves that aren't periodic.
The analysis of variance from step (1) provides us with a measure
of ``scatter'' in a light curve, after being folded into a given
period. Ideally, when the period is correct, the folded data are
neatly arranged, with minimal scatter due to noise. In contrast,
when the light curve is folded into an incorrect period, the data
are randomized and the scatter is increased. In order to quantify
this, we measure the amount of scatter in each of the tested
periods, and calculate the number of sigmas the minimum scatter is
from the mean scatter. We call this quantity, the periodic
strength score. In an attempt to minimize the number of
non-eclipsing binaries that continue to the next step, while
maximizing the number of eclipsing binaries that pass through, we
chose a minimum periodic strength score cutoff of 6.5. In addition
to this, we set a requirement that the variables' period be no
longer than 200 days, which guarantees at least four foldings.
These two criteria filtered out approximately $90\%$ of all the
light curves in OGLE~II dataset.

\begin{figure}
\epsscale{0.85}
\plotone{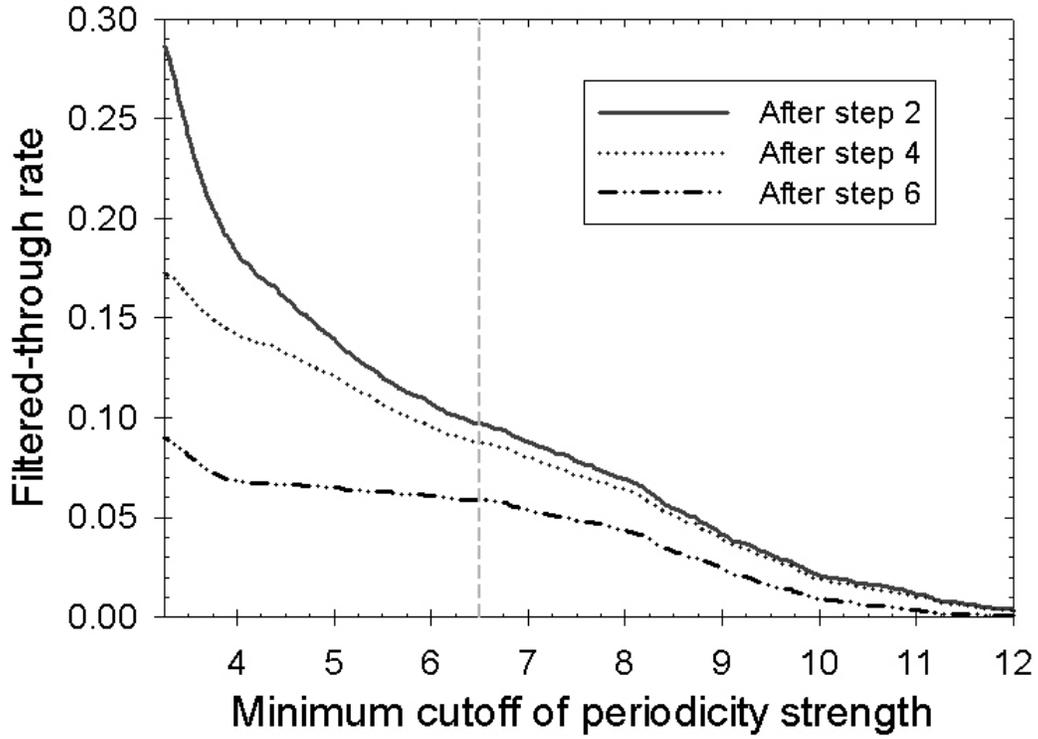}
\vspace{1in}
\caption{The pipeline filtration factions of the variables within bulge field 33 of OGLE II
 (N=4526), with varying periodicity strength cutoffs. The
vertical dashed line indicates the chosen cutoff for our pipeline ($>6.5$).\newline
\footnotesize{[See \texttt{http://cfa-www.harvard.edu/$\sim$jdevor/DEBiL.html} for a high-resolution version of this figure]}}
\label{figFilters}
\end{figure}

In order to test the effectiveness of these filtration criteria,
we measured the filtration rates for field 33, a typical OGLE~II
bulge field (fig.~\ref{figFilters}). We then repeated this
measurement for a range of periodicity strength cutoffs (the
200-day criterion remained unchanged). For periodicity strength
cutoffs up to 4, there is a sharp reduction in the number of
systems, as non-variable light curves are filtered out. Users
should be aware that constant light curves can be well-fit by
degenerate DEBiL models. By filtering out systems with such low
periodicity strength we correctly remove these systems, and in so
doing noticeably lower the total number of well-fitted systems.
Raising the filtration cutoff, up until about 6.5, will continue
reducing the filter-through rate, but with only a small impact on
the number of well-fit systems. Further raising the cutoff will
begin again reduce number of fitted systems, this time removing
good systems. We may conclude from this test that the optimal
periodicity strength cutoff is between 4 and 6.5, so that
non-variable systems are mostly filtered out while eclipsing
binaries are mostly filtered through. Since the filter-through
rate monotonically decreases as the cutoff is raised, the pipeline
becomes significantly more computationally efficient at the high
end of this range, thus bringing us to our cutoff choice of 6.5.
Users with more computing power at their disposal may consider
lowering the cutoff to the lower end of this range. In so doing
they slightly reduce the risk of filtering out eclipsing binaries,
at the price of significantly lowering the pipeline computational
efficiency.

\placefigure{figFilters}

\subsection{The Second Tier -- DEBiL Fitter}
\label{subsecDEBiLfitter}

 Steps (3) through (5) are performed within the
DEBiL program. Step (3) provides an ``initial guess'' for the
model parameters. It identifies and measures the phase, depth and
width of the two flux dips which occur in each orbit. Using a set
of equations that are based on simplified analytic solutions for
detached binary systems \citep{Danby64,MallenOrnelas03,Seager03},
DEBiL produces a starting point for the fitting optimization
procedure. Step (4) filters out light curves with out-of-bound
parameters, so to protect the following step. In practice this
step is remarkably lax. It typically filters out only a small
fraction of the light curves that pass through it, but those that
are filtered out are almost certain not to be eclipsing binaries.
Step (5) fits each light curve to the 8-parameter DEBiL model (see
\S\ref{secMotivation}), fine-tunes it, and estimates its parameter
uncertainties. This step starts with the light curve that would
have been seen with the ``initial guess'' parameters. It then
systematically varies the parameters according to an optimization
algorithm, so as to minimize the square of the residuals in an
attempt to converge to the best fit. The optimization algorithm
chosen for the DEBiL fitter is the downhill simplex method
combined with simulated annealing \citep{Nelder65,Kirkpatrick83,
Vanderbilt83,Otten89,Press92}. This algorithm was selected for its
simplicity, speed and its relatively long history at reliably
solving similar problems, which involve locating a global minimum
in a high-dimensional parameter-space. Other methods that were
considered are gradient-based \citep{Press92} and genetic
algorithms \citep{Holland92,Charbonneau95}. Gradient-based
(steepest descent) algorithms can converge very quickly to a local
minimum, but are not designed for finding the global minimum. In
addition, the difficulty in calculating the gradient of
non-analytic function slows these algorithms considerably and
causes them to be less robust. Genetic algorithms provide a
promising new approach for locating global minima, with the unique
advantage of being parallelizable. Unfortunately, their
implementations are more complicated, while not having a
significant advantage in speed or reliability over the downhill
simplex method.

Determining a convergence threshold at which to stop optimization
algorithms is known to be a hard problem \citep{Charbonneau95}.
This is because the convergence process of these algorithms will
go through fits and stops. The length of the ``stops'', whereby
the convergence does not significantly improve, becomes longer
with time, ultimately approaching infinity. Since there is no
known way to generally predict these fits and stops, we chose
simply to have the algorithm always run for 10,000 iterations (can
be adjusted at the command line). This number was found to be
adequate for OGLE light curves (see \S\ref{secTests}), with larger
numbers not showing a significant improvement in the convergence.
Using a constant number of iterations has the significant benefit
of enabling the user to make accurate predictions of the total
computing time that will be required. At the end of the 10,000
iterations, the best solution encountered so far is further
fine-tuned, so to guarantee that it is very close to the bottom of
the current minimum. In our implementation we make sure every
parameter is within $0.1\%$ of the minimum ($p_{min}$).

Finally, DEBiL attempts to estimate the uncertainties of the
fitted parameters. This is done by perturbing each parameter by a
small amount ($\Delta p$) and measuring how sensitive the model's
reduced chi square ($\chi_\nu^2$) is to that parameter. In our
implementation we set the perturbation to be $0.5\%$ of $p_{min}$.
At each parameter perturbation measurement, the remaining
parameters are re-fine-tuned\footnote{We limited the number of
iterations for this task, so that it won't become a computational
bottleneck. But since the perturbations are small, we could use a
greedy fitting algorithm, for which this is rarely a problem.}, so
to take into account the parameters' covariances. We then use a
second order Taylor expansion to derive the second derivative of
$\chi_\nu^2$ at the minimum, which is used to extrapolate the
local shape of the $\chi_\nu^2$-surface:

\begin{equation}
\chi_\nu^2 \left( p_{min} + \Delta p \right) \simeq \chi_\nu^2
\left( p_{min}  \right) + \frac{1}{2} \cdot \frac{\partial^2
\chi_\nu^2}{\partial p_{min}^2 } \cdot \left( {\Delta p } \right)^2
\end{equation}

\begin{figure}
\epsscale{0.85}
\plotone{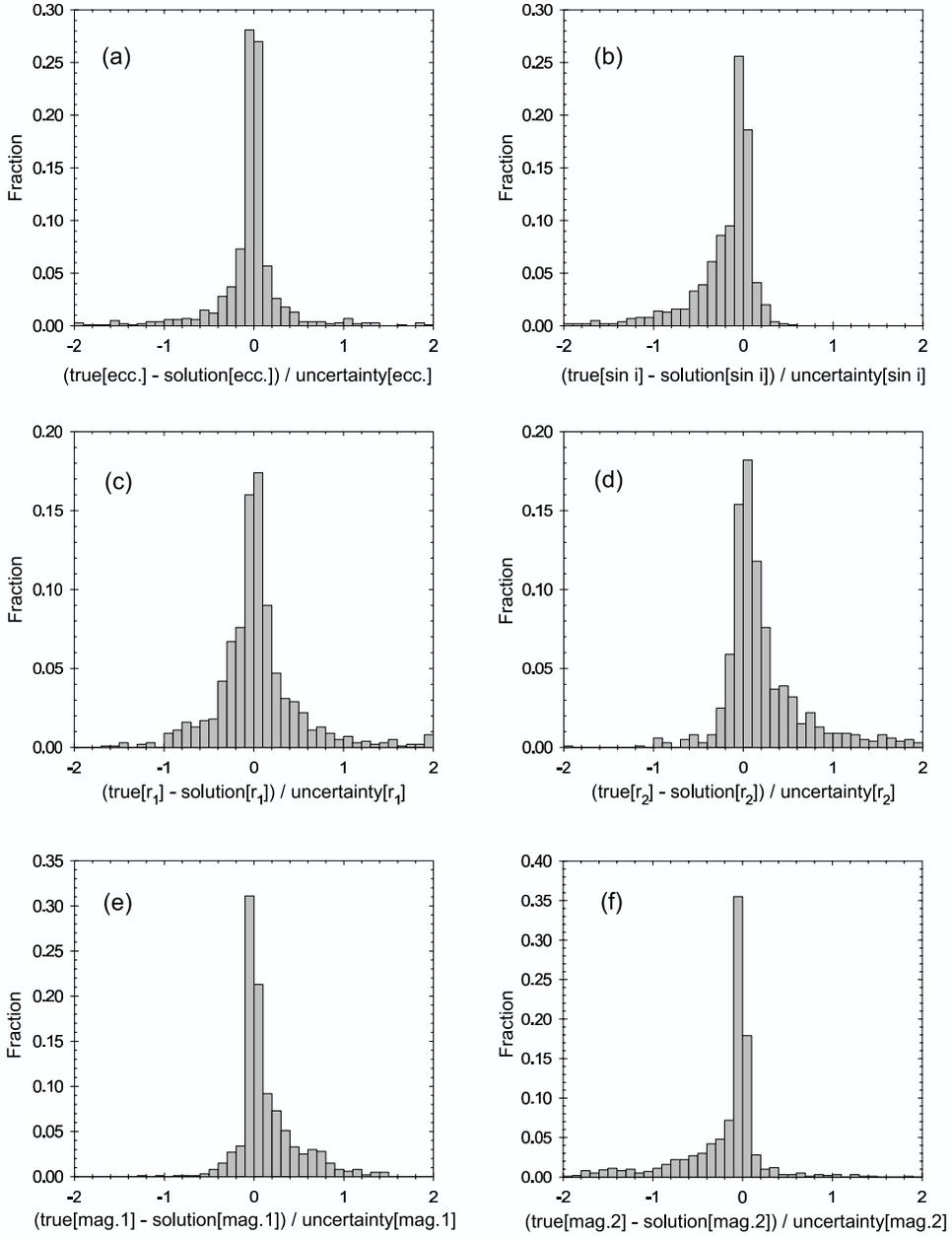}
\caption{Histograms of the error
distribution in DEBiL fitted parameters. This plot was created by
fitting 1000 simulated eclipsing light curves with 5\% Gaussian
photometric noise. The fitting error of each parameter was
normalized by its estimated uncertainty (as defined in
\S\ref{subsecDEBiLfitter}). The distributions seen here are not
Gaussian, but rather have a slender peak, and long tails (i.e.
large kurtosis). The distributions also have varying degrees of
skewness, which is discussed in \S\ref{subsecLimitations}.}
\label{figAll05Hist}
\end{figure}

We chose to employ a non-standard definition for the DEBiL
parameter uncertainties, which seems to describe the errors of all
the fitted parameters (fig.~\ref{figAll05Hist}) far better than
the standard definition \citep{Press92}. In the standard
definition, the uncertainty of a parameter is equal the size of
the perturbation from the parameter's best-fit value, which will
raise $\chi_\nu^2$ by $1/\nu$, while fitting the remaining
parameters. The reasoning behind this definition implicitly
assumes that $\chi_\nu^2$ is a smooth function. But in fact the
$\chi_\nu^2$-surface of this problem is jagged with numerous local
minima. These minima will fool the best attempts at converging to
the global minimum and cause the parameter errors to be far larger
than the standard uncertainty estimate would have us believe. For
this reason we adopted a non-standard empirical definition for the
parameter uncertainties ($\varepsilon_p$). In our variant, the
uncertainty of a parameter is equal the size of the perturbation
from the parameter's best-fit value, which will \textit{double}
$\chi_\nu^2$, while fitting the remaining parameters:

\begin{equation}
\chi_\nu^2 \left( p_{min} + \varepsilon_p \right) = 2 \chi_\nu^2
\left( p_{min} \right)
\end{equation}

This definition assigns larger uncertainties to more poorly-fit
models. In addition, it is insensitive to systematic over- or
under-estimated photometric uncertainties, which are all too
common in many light curve surveys. Using the previous two
equations, we can estimate $\varepsilon_p$ as:

\begin{equation}
\varepsilon_p  \simeq {\Delta p \cdot \sqrt {\frac{\chi_\nu ^2
\left( p_{min} \right)} {\chi_\nu ^2 \left( {p_{min} + \Delta p}
\right) - \chi_\nu^2 \left( p_{min} \right)}} }
\end{equation}

Note that the standard uncertainty is approximately:
$\varepsilon_p / \sqrt {\chi ^2}$ , so users can easily convert to
it, if so desired.

\placefigure{figAll05Hist}

Step (6) is the final gatekeeper of our pipeline. It evaluates the
model solutions and filters out all but the ``good'' models
according to some predefined criteria. DEBiL users are expected to
configure this step so to fit the particular needs of their
research. To this end, DEBiL provides a number of auxiliary tests
designed to quantify how well the model fits the data. Reduced
chi-square and fitness score values measure the overall quality of
the fit, while scatter score and waviness values measure local
systematic departures of the model from the data (see appendix A).

Additional filtering criteria are usually needed in order to
remove non-eclipsing-binary light curves that have either (a)
overestimated uncertainties that produce low reduced chi-square
results, or (b) look deceptively similar to eclipsing binary light
curves. Filtration criteria should be placed with great care in
order to minimize filtering out ``good'' light curves. In order to
handle overestimated uncertainties (a), one can filter out models
with low fitness score. Handling non-eclipsing-binary light curves
that look like binary light curves (b) is considerably more
difficult. Many of these problematic light curves are created by
pulsating stars (e.g. RR-Lyrae type~C), which have sinusoidal
light curves that resemble those of contact binaries. To this end,
DEBiL also provides the reduced chi-square of a best-fit
sinusoidal function of each light curve. If this value is similar
or lower than the model's reduced chi-square, then is it likely
that the light curve indeed belongs to a pulsating star.

\subsection{Limitations}
\label{subsecLimitations}

In the previous subsection we discussed the considerable
difficulties in finding the global minimum in the jagged structure
of the $\chi_\nu^2$-surface. At this point we must further add,
that since the data is noisy and the model is imperfect, the true
solution might not be at the global minimum. For the lack of
better information, we can only use the global minimum as the
point in parameter-space that is the most \textit{likely} to be
the true solution.

Another source of errors are systematic fitting biases, which must
especially be taken into account when making detailed population
studies. Two main sources of these biases are imperfect models and
asymmetric $\chi_\nu^2$-minima. Almost all models are imperfect,
but when effects not included in the model become significant, the
optimization algorithm will often try to compensate for this by
erroneously skewing some of the parameters within the model. An
example of this is seen in semidetached binaries. The tidal
distortions of these stars are not modeled by DEBiL, so as a
result DEBiL will compensate for this by overestimating their
radii. Most model imperfections are flagged by a large reduced
chi-square. Surprisingly, for tidal distortions, the reduced
chi-square is not significantly increased. For this reason we
provide a ``detached system'' criterion that will be described in
\S\ref{secResults}. The effects of asymmetric $\chi_\nu^2$-minima
are more subtle. In such cases a perturbation to one side of a
$\chi_\nu^2$-minimum raises $\chi_\nu^2$ less than a perturbation
to the other side. Thus random noise in the data will cause the
parameters to be systematically shifted more often in the former
direction than in the latter. We believe that these biases, which
are universal to all fitting programs, can be corrected after the
fact. We chose not to do this in order to avoid having to insert
any fudge factors into DEBiL. But we acknowledge that this may be
necessary in order to extend the regimes in which DEBiL is
reliable without significantly reducing its speed.

Possibly the most problematic fitting errors are those that are
caused by mistaken light curve periods. For eclipsing binaries
with strong periodicity scores, the main cause of this is the
confusion between two types of light curves: (a) light curves with
very similar and equally-spaced eclipses, and (b) light curves
with an undetected eclipse. Both these types of light curves will
appear to have a single eclipse in their phased light curve. Light
curves with similar eclipses (a) will cause the period finder to
return a period that is half the correct value\footnote{Strictly
speaking, this is not an error on the part of the period finder,
since it is in fact returning the best period from its
standpoint.}, folding the primary and secondary eclipse over one
another. In contrast to this, light curves with an undetected
eclipse (b), either because it's hidden within the noise or
because it's in a phase coverage gap, will have the correct
period. Because the undetected eclipse is necessary for
determining a number of the fitting parameter, we are not able to
model this type of light curve. Fortunately, light curves with
similar eclipses (a) can be easily modeled by simply doubling
their period. Since we can't generally distinguish between these
two types of light curves, the DEBiL fitter treats all the
single-eclipse light curves as type (a), doubling their period and
fitting them as best it can. Whenever such a period-doubling
occurs, DEBiL inserts a warning message into the log file. These
light curves should be used with increased scrutiny. This problem
is further compounded in surveys, such as the OGLE II bulge
fields, which consist of many non-eclipsing-binary variables. As
mentioned in \S\ref{subsecDEBiLfitter}, some of these systems look
deceptively similar to eclipsing binaries. Since their brightness
oscillation typically consists of a single minima, they too will
have their period doubled. In conclusion, unless the systems with
doubled periods are filtered out, there will be an erroneous
excess of light curves with similar primary and secondary
eclipses. In turn, this will manifests itself in an excess of
model solutions with stars of approximately equal surface
brightness.

Even though our discussions of possible causes and remedies for
limitations stemmed from our experience with our particular pipeline,
many of these points are also likely to apply to other similar
pipelines and fitting procedures.

\section{Tests}
\label{secTests}

\begin{figure}
\epsscale{0.85}
\plotone{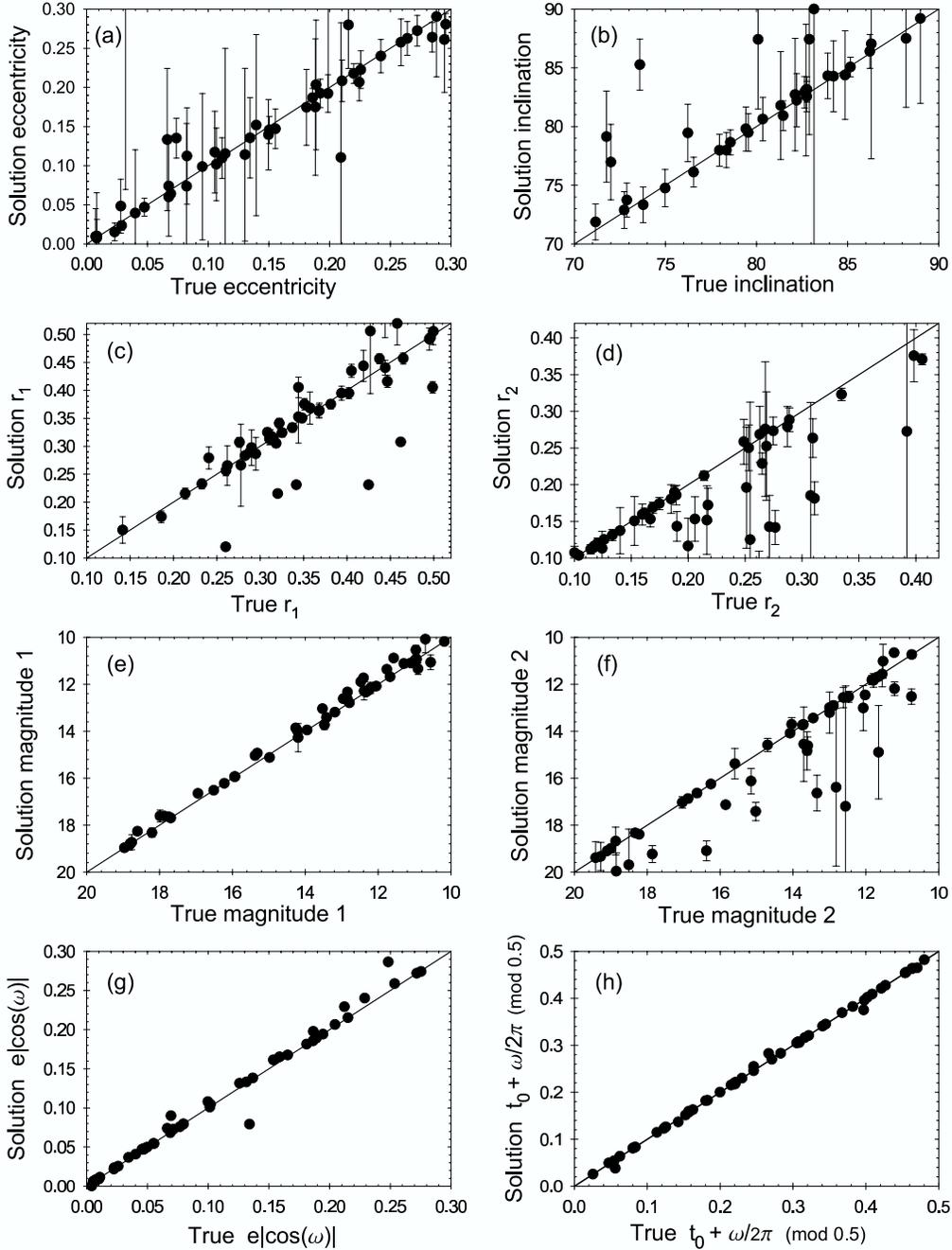}
\caption{The fitted solutions
versus the true solution in 50 simulated eclipsing binary light
curves with Gaussian photometric noise of 1\%. We simulated a
uniform distribution of parameters, with the only requirement
being that both eclipse dips were detectable through the noise.
Panels (a) through (f) show the fits of DEBiL model parameters
with their uncertainties, as defined in \S\ref{subsecDEBiLfitter}.
Panels (g) and (h) combine parameters so that they describe
prominent features of the light curve (respectively, the
separation and offset of the eclipse centers). In these
combination, the parameter errors largely cancel out, so that the
formal uncertainties should not be used.}
\label{figAll01}
\end{figure}

In order to test the DEBiL fitter, we ran it both on simulated
light curves and published, fully analyzed, observed light curves
\citep{Lacy00,Lacy02,Lacy03}. Figure~\ref{figAll05Hist} shows the
results of fitting 1000 simulated light curve, with $5\%$ Gaussian
photometric noise. Figure~\ref{figAll01} provides a more detailed
look at the fits to 50 simulated light curves, with $1\%$ Gaussian
photometric noise, giving results comparable to those of
\citet{Wyithe01}. Not surprisingly, when less noise was inserted
into the light curve, the fitter estimated considerably smaller
uncertainties.

\begin{figure}
\plotone{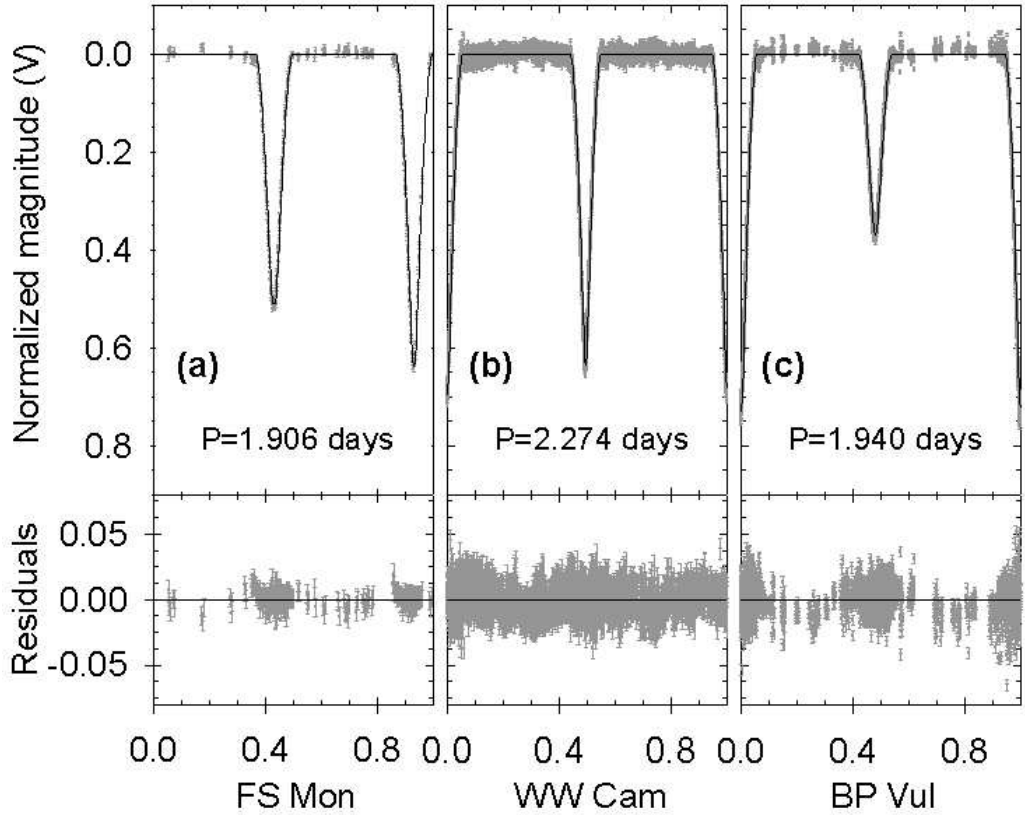}
\caption{The phased light curves with the DEBiL model best fit (solid line) and its residuals,
for the eclipsing binary systems: (a) FS Monocerotis; (b) WW Camelopardalis; (c) BP Vulpeculae.\newline
\footnotesize{[See \texttt{http://cfa-www.harvard.edu/$\sim$jdevor/DEBiL.html} for a high-resolution version of this figure]}}
\label{figLacy}
\end{figure}

While the simulated light curves are easy to produce and have
known parameter values, observed light curves are the only ones
that can provide a true reality check. To this end, we also
reanalyze three published light curves (fig.~\ref{figLacy}):

\begin{itemize}
\item Table~\ref{tableLacyA}: FS Monocerotis \citep{Lacy00}
\item Table~\ref{tableLacyB}: WW Camelopardalis \citep{Lacy02}
\item Table~\ref{tableLacyC}: BP Vulpeculae \citep{Lacy03}
\end{itemize}

We present here a comparison between the aforementioned published
photometric fits, using the Nelson-Davis-Etzel model implemented
by EBOP \citep{Etzel81,Popper81b} and the DEBiL fits. For all
three cases, we used the V-band observational data and set DEBiL's
limb darkening quadratic coefficients to the solar V-band values
\citep{Claret03}. We found that when applying the physically
correct limb darkening coefficients, the improvements in the
best-fit model were negligible compared to the uncertainties. For
this reason, we chose to use solar limb darkening coefficients
throughout this project.

\placefigure{figAll01}
\placefigure{figLacy}
\placetable{tableLacyA}
\placetable{tableLacyB}
\placetable{tableLacyC}

\section{Results}
\label{secResults}

 We used the aforementioned pipeline to identify
and analyze the eclipsing binary systems within the bulge fields
of OGLE~II \citep{Udalski97,Wozniak02}. The final result of our
pipeline contained only about 5{\%} of the total number of light
curves we started with. The filtration process progressed as
follows:

\begin{itemize}
\item Total number of OGLE~II (bulge fields) variables: 218699
\item After step (2), with strong periodicity and periods of 0.1-200 days: 19264
\item After step (4), the output of the DEBiL program: 17767
\item After step (6), with acceptable fits to binary models ($\chi_\nu^2 <4$): 10862
\end{itemize}

\begin{figure}
\plottwo{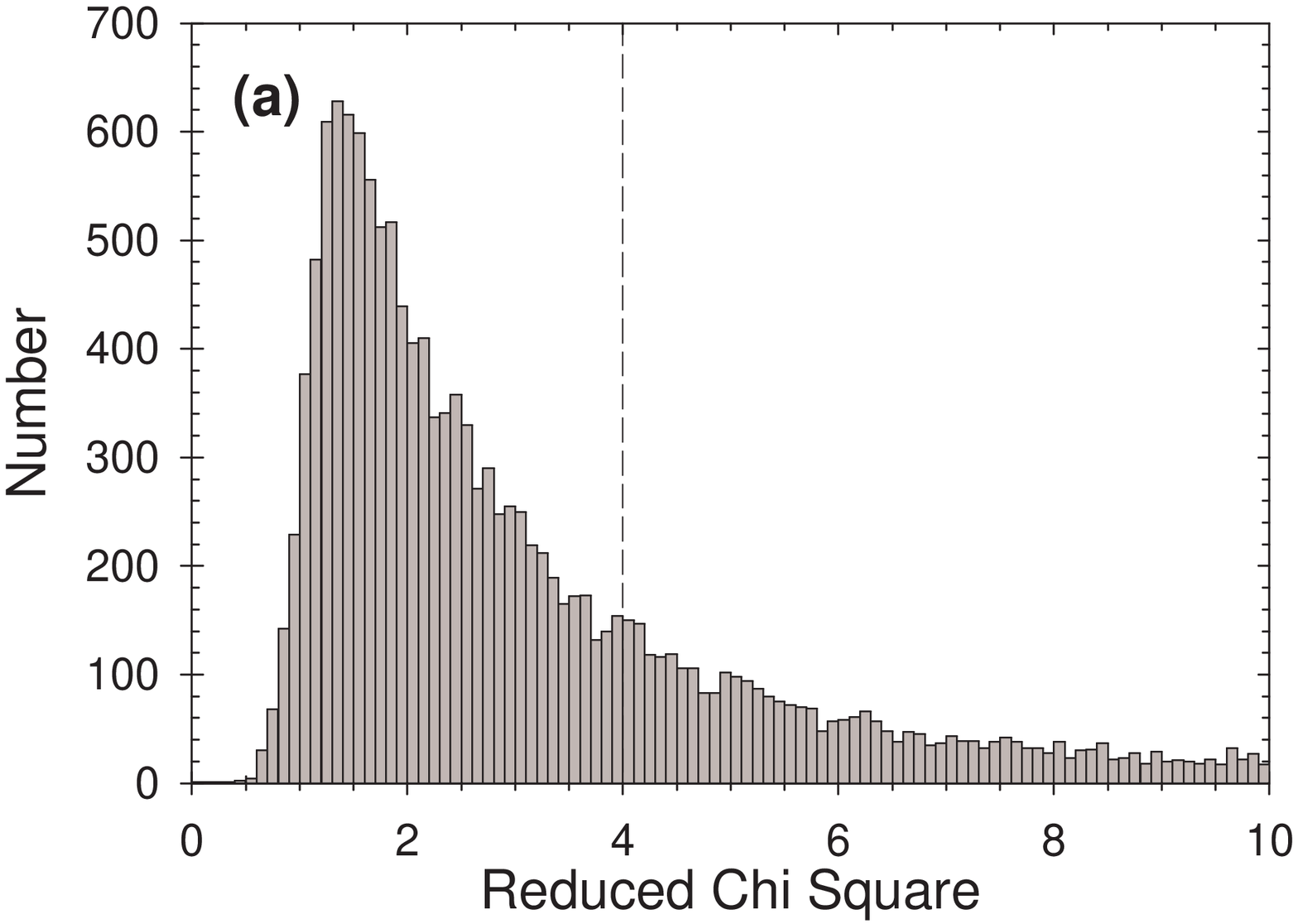}{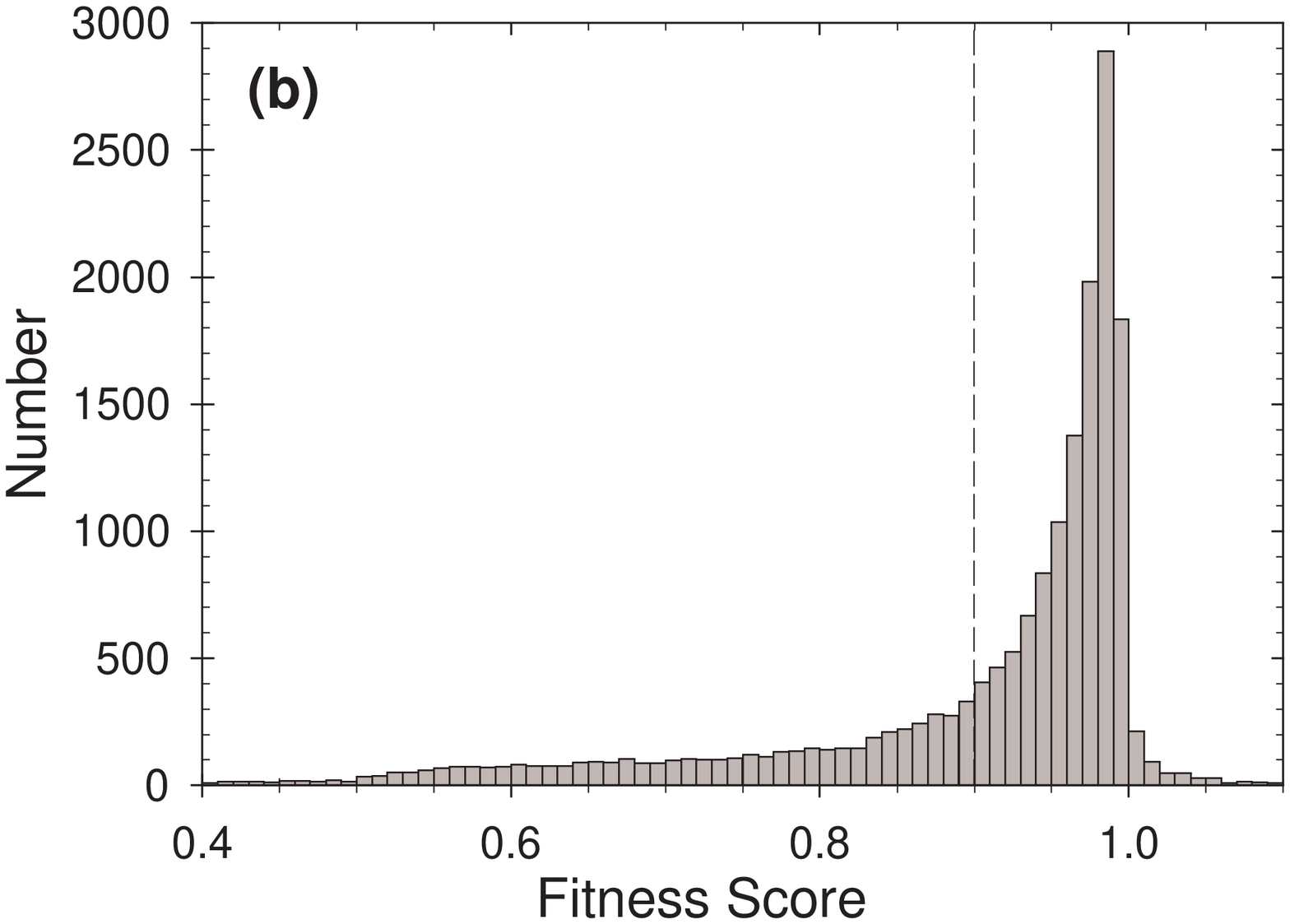}
\caption{The distribution of: (a)
reduced chi-square ($\chi_\nu^2$), and (b) fitness score results
of the DEBiL model fits for OGLE~II bulge (see appendix A). The
vertical dashed lines mark the filtration thresholds used in our
pipeline ($<4$ and $>0.9$ respectively). Both tests show a
definite peak near 1, indicating that it is more likely that
DEBiL will produce a ``good'' fit than a ``bad'' fit.}
\label{figHistChi}
\end{figure}

\begin{figure}
\plotone{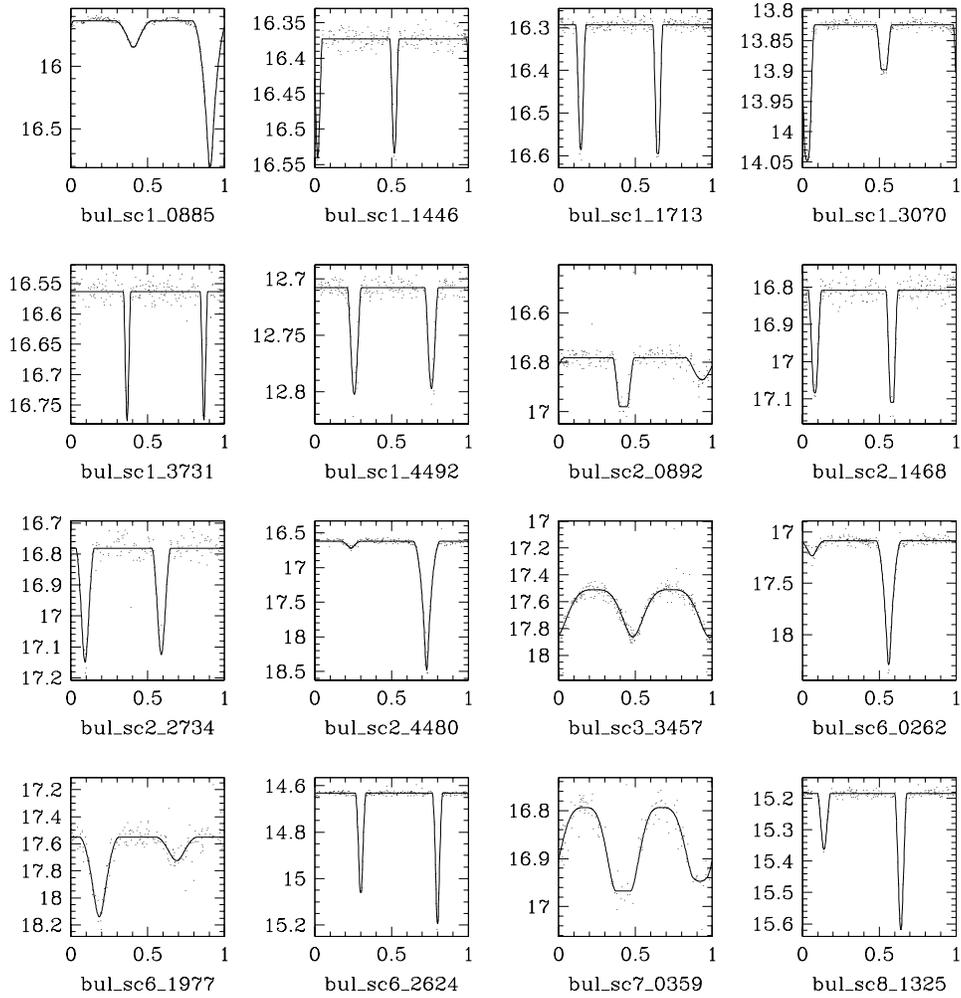}
\caption{Selected examples of OGLE~II bulge field light curves, with their DEBiL fits.}
\label{figCatalog}
\end{figure}

Most of the fits that reach step (6) can be considered successful
(fig.~\ref{figHistChi}). It is then up to the user to choose
criteria for light curves that are of interest, and to define a
threshold for the quality of the fits. In our pipeline we chose a
very liberal quality threshold ($\chi_\nu^2 <4$), so to allow
through light curves with photometric uncertainties that are too
small (a common occurrence), and to leave users with a large
amount of flexibility in their further filtrations. We list the
first 15 DEBiL fits that passed though step (6) in
table~\ref{tableDEBiL}. Figure~\ref{figCatalog} shows another
sampling of models, with their corresponding phased light curves.
The complete dataset of OGLE II bulge models, both plotted and in
machine readable form, is available online.

Since the filter at step (6) may not be stringent enough for many
application, we also provide our results after each of two further
levels of filtration:

\begin{itemize}
\item Non-pulsating (fitness score $> 0.9$; non-sinusoidal light curves): 8471
\item Detached systems (both stars are within their Roche limit): 3170
\end{itemize}

For non-sinusoidal light curves, we require that the DEBiL fit
have a smaller reduced chi square than the best-fit sinusoidal
model. For the Roche limit calculation, we assumed an early main
sequence mass-radius power law relation: $R \propto M^{0.652}\;$
\citep{Gorda98} and set it in a third order approximation of the
Roche radius \citep{deLoore92}. The resulting approximations are:

\begin{eqnarray}
R_{Roche,1}/a & \simeq & 0.37771+0.31054x+0.04324x^2+0.08208x^3\\
R_{Roche,2}/a & \simeq &
\cases{0.37710-0.32684x-0.01882x^2-0.023812x^3, & if x $\ge 0.65$
\cr 0.37771-0.31054x+0.04324x^2-0.08208x^3, &otherwise\cr}
\end{eqnarray}

Where: $x\equiv \log \left( {R_1 /R_2 } \right)\;$, assuming: $R_1
\ge R_2$.

\placefigure{figHistChi}
\placefigure{figCatalog}
\placetable{tableDEBiL}

\begin{figure}
\plotone{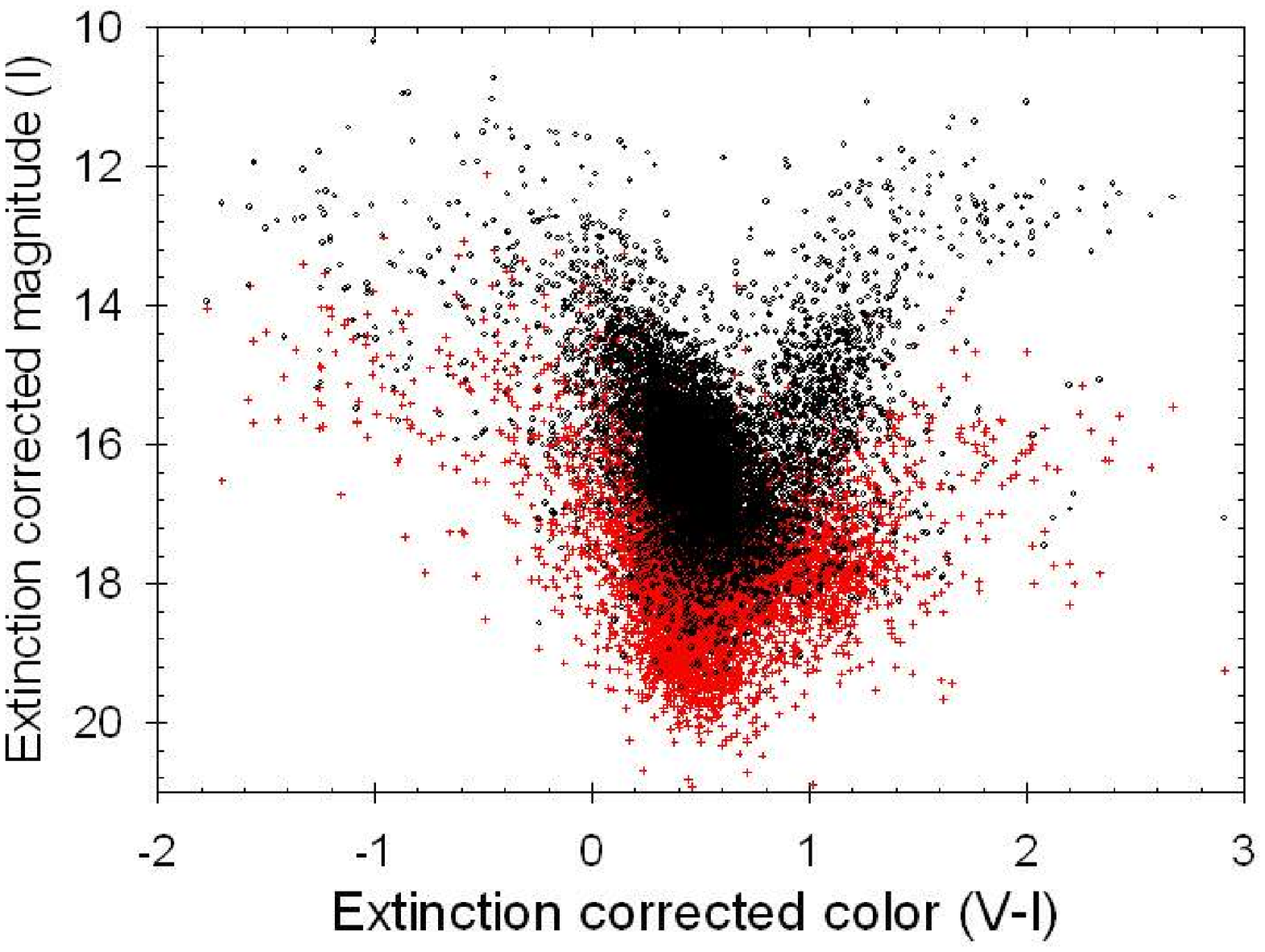}
\caption{The color-magnitude diagram of both the primary (circles) and secondary (crosses) stars of the DEBiL models.
The color of each star is the combined color of the binary, from \citep{Udalski02}.\newline
\footnotesize{[See \texttt{http://cfa-www.harvard.edu/$\sim$jdevor/DEBiL.html} for a high-resolution version of this figure]}}
\label{figColorMag}
\end{figure}

\begin{figure}
\plotone{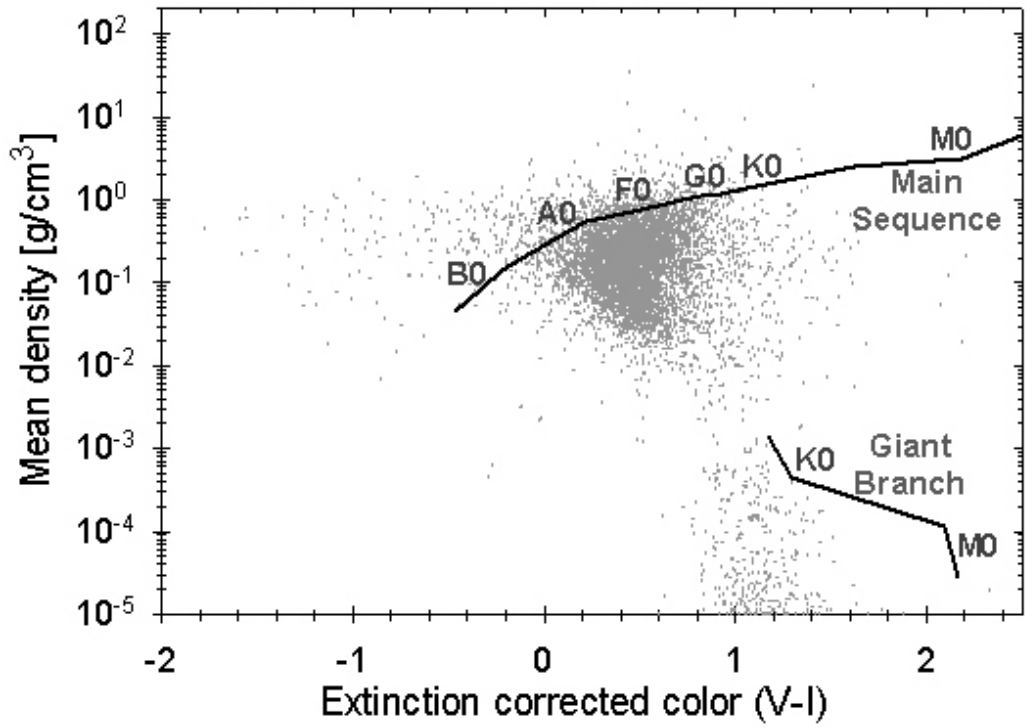}
\vspace{1in}
\caption{Color-density diagram of DEBiL models. The solid lines
trace the main sequence stars and giants \citep{Cox00}. Notice the
strong observational selection bias for main sequence F-type and
G-type stars in the OGLE~II bulge fields.\newline
\footnotesize{[See \texttt{http://cfa-www.harvard.edu/$\sim$jdevor/DEBiL.html} for a high-resolution version of this figure]}}
\label{figColorDens}
\end{figure}

In order to better interpret the data derived by the DEBiL
pipeline, we combined it with color (V-I) and magnitude
information \citep{Udalski02}, as well as an extinction map of the
galactic bulge \citep{Sumi04}. Thus for each eclipsing system, we
also have its extinction corrected combined I-band magnitude and
V-I color. Since many of the stars in the bulge fields are not in
the galactic bulge but rather in the foreground, it is likely they
will be overcorrected, making them too blue and too bright. These
stars can be seen in both the color-magnitude diagram
(fig.~\ref{figColorMag}) and in the color-density diagram
(fig.~\ref{figColorDens}). With this qualification in mind, the
color-density diagram provides a distance independent tool for
identifying star types. Because the measured values result from a
combination of the two stars in the binary, the values aren't
expected to precisely match either one of the stars in a binary.
Remarkably, using the max density measure instead of the mean
density (appendix B) this problem seems to be considerably
lessened.

\subsection{Population Distributions}

Due to the limitations of the OGLE observations, analysis and
subsequent filtrations, there are a myriad of complex selection
effects that need to be accounted for. For this reason, we
hesitate to make any definite population statements in this paper,
although there are a number of suggestive clusterings and trends
that merit further scrutiny.

\begin{figure}
\plotone{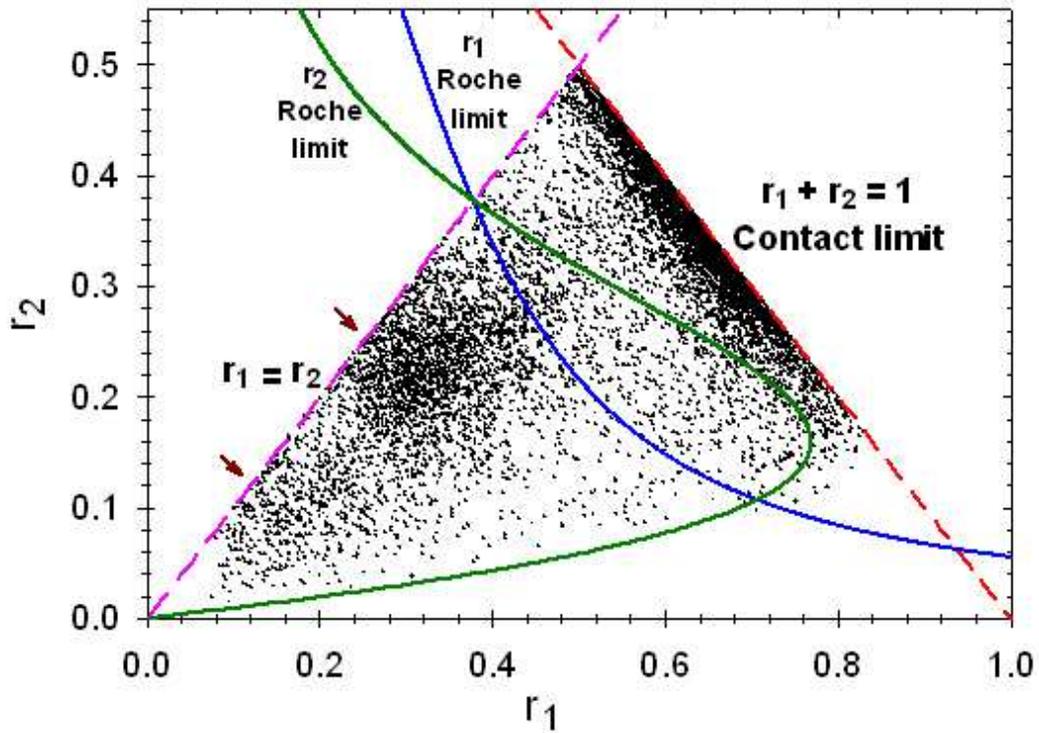}
\vspace{1in}
\caption{Radius-radius plot. The dashed lines
mark the outer limit of the data. The left side is bounded by the
fact that by definition: $r_1 \geq r_2$ , while the right side is
bounded by the physical contact limit of the stars. The two arrows
mark anomalous clusterings. The two solid curves approximate the
location where the primary and secondary stars reach their
respective Roche limit (eq. 4 \& 5). Systems between the two solid
curves are semidetached, systems to their left are detached, and
systems to their right are contact or overcontact systems.\newline
\footnotesize{[See \texttt{http://cfa-www.harvard.edu/$\sim$jdevor/DEBiL.html} for a high-resolution version of this figure]}}
\label{figR1R2}
\end{figure}

When considering the distribution of $r_{1,2} \equiv R_{1,2} / a$
for eclipsing binary systems, one would expect a comparably smooth
distribution as determined by the binary star IMF, binary orbit
dynamics and observational/detection selection effects. This
distribution can be best seen in a radius-radius plot
(fig.~\ref{figR1R2}). This plot has a few features which merit
discussion. One feature is that the number of systems rapidly
dwindles as their radii become smaller. This should not be
surprising as selection effects dominate the expected number of
observed systems, for small radii (this point will be elaborated
at the end of this subsection). A second, far more surprising
feature, is the appearance of three clusterings: along the contact
limit, around $\left(r_1 = 0.33, r_2 = 0.23 \right)$ and around
$\left(r_1 = 0.15, r_2 = 0.09 \right)$. These three clusterings
are most likely artificial, caused by two technical limitations
which are discussed in \S\ref{subsecLimitations}. The first two
clusterings were likely formed when many of the semidetached
systems that populated the region between the clusterings, were
swept into the contact limit, because their radii were
overestimated. The third clustering, though less pronounced, seems
to echo the structure of the second clustering, only with about
half its radii. This hints at the possibility that the periods of
some of these systems were doubled when they shouldn't have been,
probably because of an undetected eclipse. It is worth mentioning
that both the second and third clusterings are centered at a
significant distance from $r_1 = r_2$, which is where the clusters
would have been located, if $r_1$ and $r_2$ had been independent
variables.

\begin{figure}
\plotone{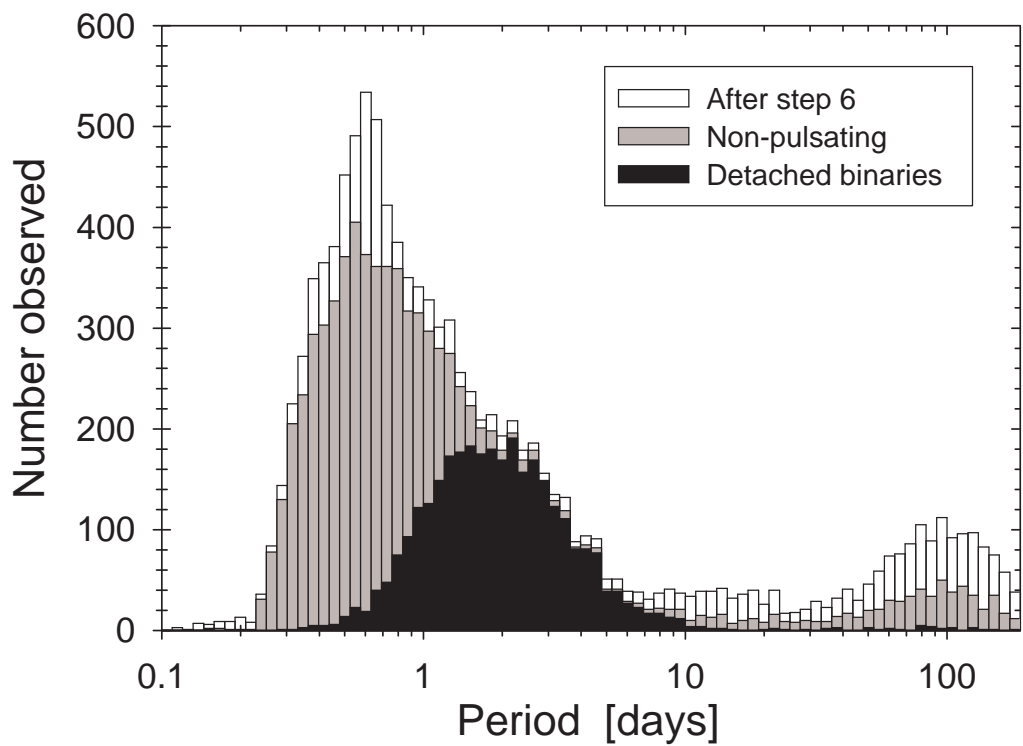}
\caption{The period distribution of OGLE~II bulge eclipsing binaries, following various stages of filtration.}
\label{figPeriod1}
\end{figure}

\begin{figure}
\plottwo{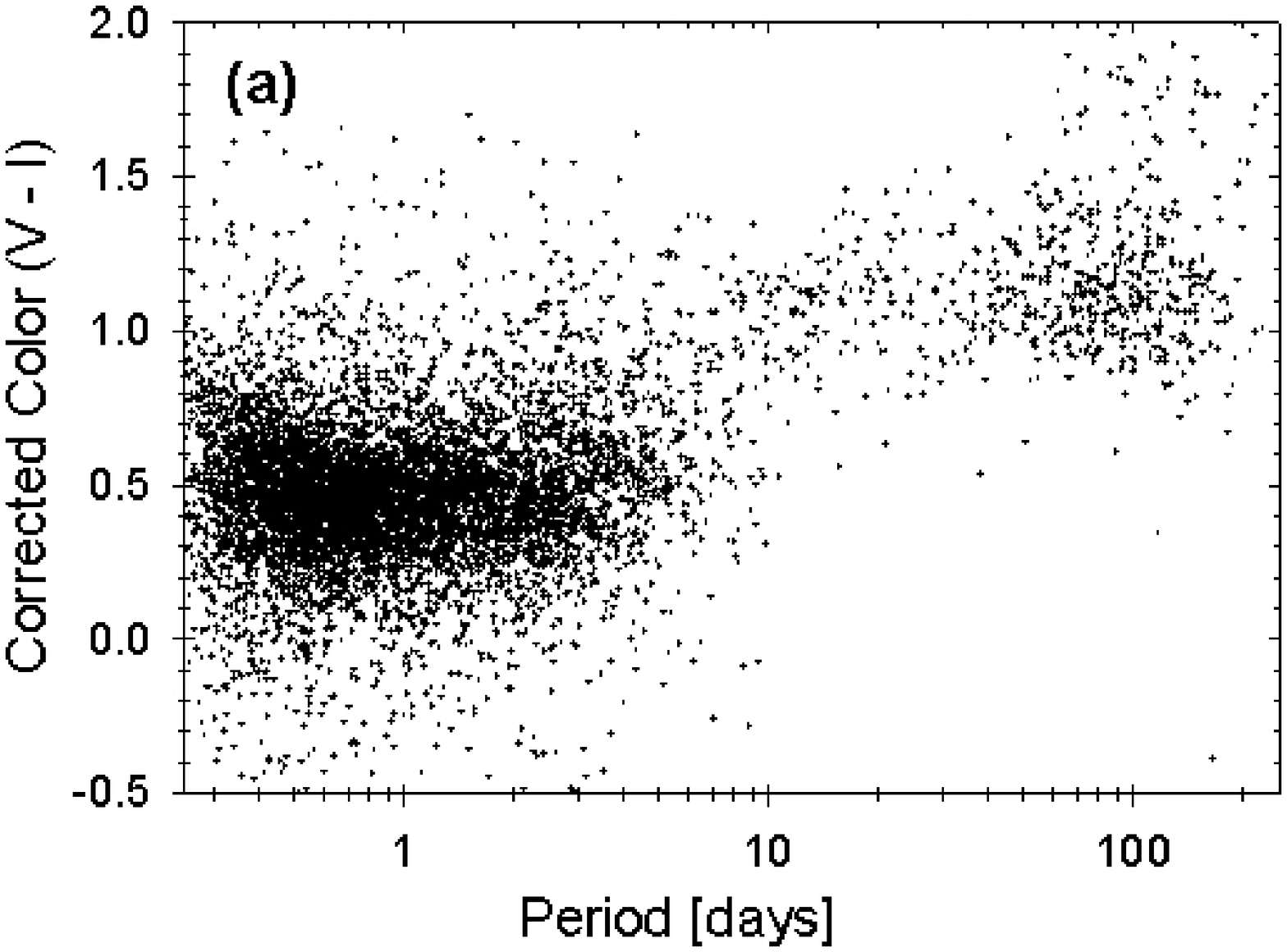}{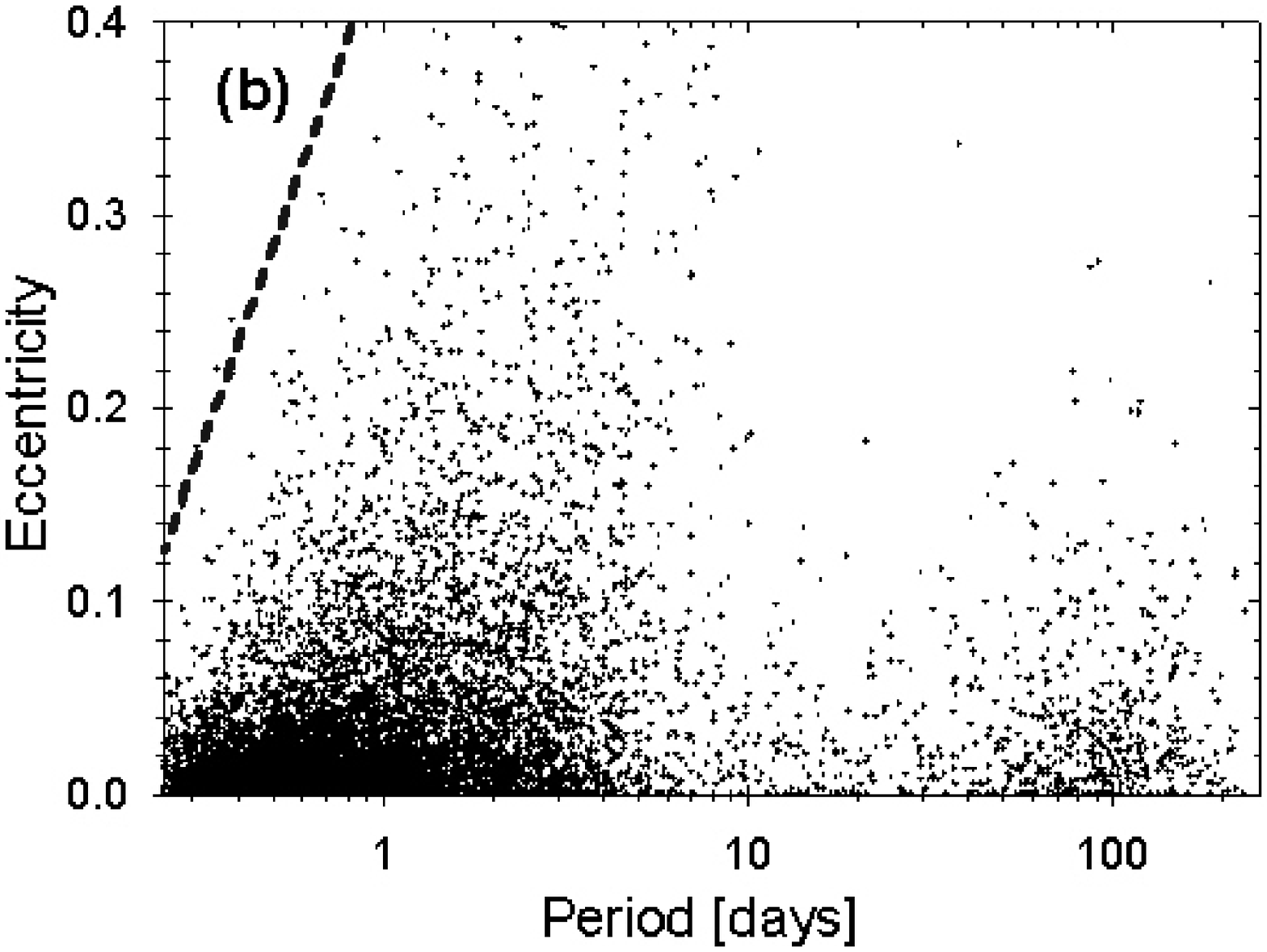}
\caption{(a) Color-period plot and (b) eccentricity-period plot. The upper
limit on the eccentricities of short period binaries (dashed line) is
probably due to tidal circularization.\newline
\footnotesize{[See \texttt{http://cfa-www.harvard.edu/$\sim$jdevor/DEBiL.html} for a high-resolution version of these figures]}}
\label{figEccColorPeriod}
\end{figure}

Perhaps even more surprising is the period distribution
(fig.~\ref{figPeriod1}). Using only the default filters of our
pipeline, this distribution is bimodal, peaking at approximately 1
and 100 days, and with a desert around a 20-day period. To
understand this phenomenon one should look at the scatter plots of
figure \ref{figEccColorPeriod}. In this figure we can see that
long period binaries (period $> 20$ days) are significantly redder
than average and have low eccentricities. This is possibly due to
red giants, which cannot be in short period systems. This
possibility is problematic since it contradicts the unimodal
results of previous period distribution studies
\citep{Farinella79,Antonello80,Duquennoy91}. In addition,
figure~\ref{figPeriod1} shows that with additional filtrations,
the grouping of systems around the 100-day period is greatly
reduced. All this leads us to conclude that the $\sim$100 day
period peak is probably erroneous, created by a population
contamination of pulsating stars, probably mostly semi-regulars,
which can easily be confused with contact binaries. A similar
phenomenon is seen in the spike in the number of system around a
period of 0.6 days. This increase is probably also due to
pulsating stars, only this time they are likely RR-Lyrae. They too
were largely filtered out using the techniques described in the
previous section.

Finally, we unraveled the geometric selection effects by weighting
each eclipsing light curve by the inverse of the probability of
observing it as eclipsing. For example, if with a random
orientation there is a $1\%$ chance of a given binary systems
being seen as eclipsing, we will give it a weight of 100. In this
way we tabulated each observed occurrence as representing 100 such
binary systems, the remaining of which exist but aren't seen
eclipsing and so are not included in our sample of variable stars.
A binary system with a circular orbit (e = 0) will eclipse when
its inclination angle (i) obeys: $\cos (i) < (R_1 + R_2)/a $. If
the orbital orientations are randomly distributed, then the
inclination angles are distributed as: $p(i)\propto \left| \sin
(i) \right|$. Therefore, the probability that such a binary system
will eclipse becomes:

\begin{equation}
p_{eclipse}(R_{1,2}/a,e=0) = \frac{R_1 + R_2}{a}
\end{equation}

\begin{figure}
\plotone{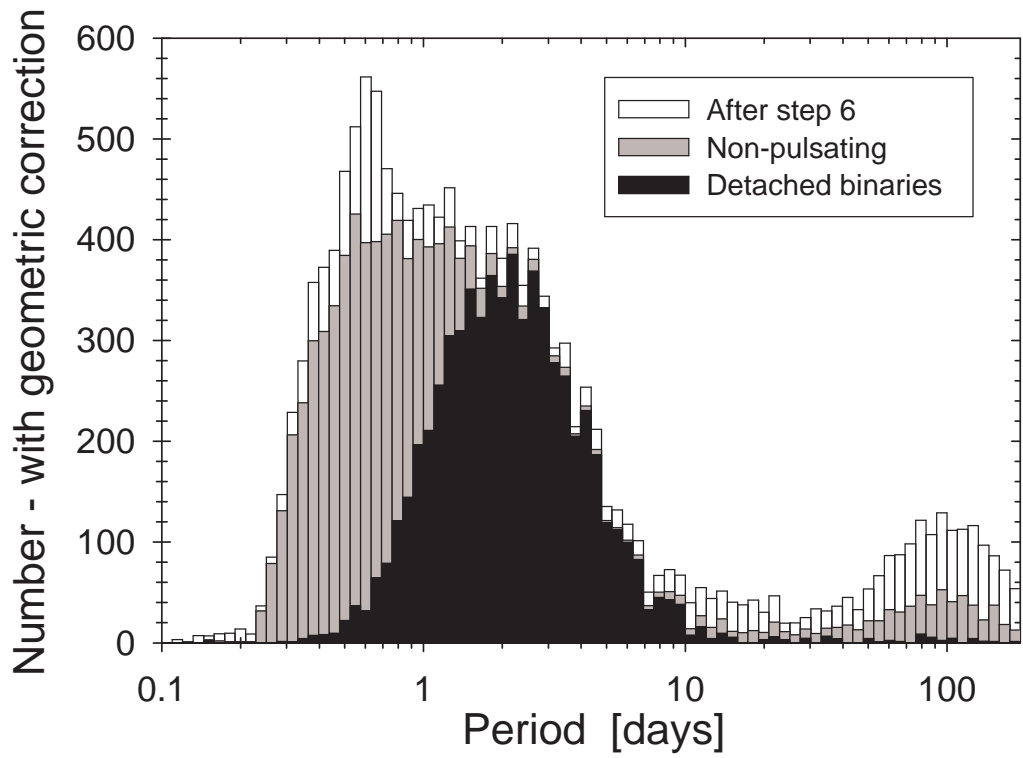}
\caption{The period distribution of OGLE~II bulge eclipsing binaries,
following various stages of filtration, after correcting for their geometric selection effect.}
\label{figPeriod2}
\end{figure}

The probability of eccentric systems (e $> 0$) eclipsing is more
difficult to calculate. We used a Monte-Carlo approach to
calculate a 1000$\times$1000 $[(R_1 + R_2)/a , e]$ probability
table, which was then used to interpolated the probability of each
fitted light curve. When applying this correction for the
geometric selection effect (fig.~\ref{figPeriod2}) we see that, as
expected, it primarily effects the detached binaries. The period
(P) distribution of detached systems, both before ($p_{eclipse}$)
and after ($p_{all}$) the geometric correction are remarkably
similar, and can both be well fit to a log-normal distribution:

\begin{equation}
p_{eclipse,all}(P) = \frac{1}{\sqrt{2\pi} \sigma P} \exp
\left({-\frac{\ln^2(P/P_0)}{2 \sigma^2}} \right)
\end{equation}

Before the geometric correction ($p_{eclipse}$) we get a fit of
$P_0 = 1.83 \pm 0.01$ days and $\sigma = 0.593 \pm 0.005$ ($r^2
\simeq 0.9926$). After the correction ($p_{all}$) we get a fit of
$P_0 = 2.06 \pm 0.01$ days and $\sigma = 0.621 \pm 0.005$ ($r^2
\simeq 0.9928$). Such log-normal distributions are generally
indicative of many independent multiplicative processes taking
place. Possibly, both in the formation of binary systems as well
as in their observational/detection selection effects.

One should be very careful when using this method to calculate the
total number of binaries. Simply comparing the number of binaries
before and after the geometric correction will result in the
conclusion that we are observing as eclipsing binaries about half
of the total number of binaries. This fraction is biased high due
to the fact that the OGLE~II survey cannot detect long period
eclipsing binaries.

\placefigure{figColorMag}
\placefigure{figColorDens}
\placefigure{figR1R2}
\placefigure{figPeriod1}
\placefigure{figEccColorPeriod}
\placefigure{figPeriod2}

\subsection{Extreme Systems}

In the previous subsection we considered the way large sets of
eclipsing systems are distributed, now we will consider individual
systems. As examples we chose to locate extreme binary systems,
systems with a parameter well outside the normal range. For this
to be done properly, we need to take great care in avoiding the
pitfalls that arise from parameter estimation errors, both
systematic and non-systematic. One pitfall is that some of the
extreme systems may have large systematic errors since they will
contain additional phenomena which the fitted model neglects.
Another, perhaps more problematic pitfall, is that we will
retrieve non-extreme systems with large errors that happen to
shift the parameter in question beyond our filtering criterion
threshold. Some examples of possible sources of large errors are
inaccuracies in the correction for dust reddening (see the
beginning of this section) and difficulties in the estimation of
the argument of periastron, which directly affects the
determination of binary's orbital eccentricity \citep{Etzel91}.
Because of these pitfalls, we expect that the human eye will be
required as the final decision maker for this task in the
foreseeable future.

We present candidates of the most extreme eclipsing systems within
the OGLE~II bulge field dataset in five categories
(table~\ref{tableExtreme}). We chose examples for which we were
comparably confident, though we would need to follow them up with
spectroscopic measurements to be certain of their designations.

\placetable{tableExtreme}

In the case of the high density, low density and blue systems the
measured value of the characteristic is a weighted mean of the two
stars in the binary system. The weighting of the high and low
density systems is described in appendix B. The weighting of the
blue systems is determined, approximately, by the bolometric
luminosity of each of the stars in the binary system.

\section{Conclusions}

We present a new multi-tiered method for systematically analyzing
eclipsing binary systems within large-scale light curve surveys.
In order to implement this method, we have developed the DEBiL
fitter, a program designed to rapidly fit large number of light
curves to a simplified detached eclipsing binary model. Using the
results of DEBiL one can select small subsets of light curves for
further follow-up. Applying this approach, we have analyzed 218699
light curves from the bulge fields of OGLE~II, resulting in 10862
model fits. From these fits we identified unexpected patterns in
their parameter distribution. These patterns are likely caused by
selection effects and/or biases in the fitting program. The DEBiL
model was designed to fit only fully detached systems, so users
should use fits of semidetached and contact binary systems with
caution. One can probably find a corrections for the parameters of
these systems, though it is best to refit them using more complex
models, which also take into account mutual reflection and tidal
effects. Even so, the DEBiL fitted parameters will likely prove
useful for quickly previewing large datasets, classification,
fitting detached systems and providing an initial guess for more
complex model fitters.

\acknowledgments

We would like to thank Robert Noyes, Krzysztof Stanek, Dimitar
Sasselov and Guillermo Torres for many useful discussions and
critiques. In addition we would like to thank Takahiro Sumi and
the OGLE collaboration for providing us with the data used in this
paper. Finally, we would like to thank Lisa Bergman for both her
editorial help and utmost support throughout this project. This
work was supported in part by NASA grants NAG5-10854 and
NNG04GN74G.

\appendix

\section{Statistical Tests}

\subsection{Fitness Score}

One of the problems with using the reduced chi-square test is that
the light curve uncertainties may be systematically overestimated
(underestimated), causing the reduced chi-square to be too small
(large). An easy way to get around this problem is by comparing
the reduced chi-square of the DEBiL model being considered, with
the reduced chi-square of an alternative model. We used two simple
alternative models:

- A constant, set to the average amplitude of the data.

- A smoothed spline, derived from a second order polynomial fit
within a sliding kernel\footnote{In our implementation, we used a
rectangular kernel whose width varies so to cover a constant
number of data points. This is needed to robustly handle sparsely
sampled regions of the phased light curve.} over the phased light
curve.

The constant model should have a larger reduced chi-square than
the best-fit model, while the spline model should usually have a
smaller reduced chi-square. In a way similar to an F-test, we
define the fitness score as:

\begin{equation}
\mbox{Fitness\ Score} \equiv \frac{\chi_\nu^2 (\mbox{const}) - \chi_\nu^2
(\mbox{DEBiL})}{\chi_\nu^2 (\mbox{const}) - \chi_\nu^2 (\mbox{spline})}
\end{equation}

This definition is useful since gross over- or underestimates of
the uncertainties will largely cancel out. Light curves that reach
step 6, will mostly have fitness scores between 0 and 1. If the
reduced chi-square of the DEBiL model equals the constant model's
reduced chi-square, the fitness score will be 0, and if it equals
the spline model's reduced chi-square, the fitness score will be
1. The fact that most of the DEBiL models have fitness scores
close to 1, and sometimes even surpassing it
(fig.~\ref{figHistChi}), provides a validation for the fitting
algorithm used.

\subsection{Scatter Score}

This test quantifies the systematic scatter of data above or below
the model, using the correlation between neighboring residuals.
The purpose of this test is to quantify the quality of the model
fit independently of the reduced chi-square test. While the
reduced chi-square test considers the amplitude of the residuals,
the scatter score considers their distribution. The scatter score
is defined after folding the n data points into a phase curve:

\begin{equation}
\mbox{Scatter\ Score} \equiv \frac{\Delta X_n \cdot \Delta X_1
+\sum\limits_{i=2}^n {\Delta X_{i-1} \cdot \Delta X_i }
}{\sum\limits_{i=1}^n {\Delta X_i^2 } }=\frac{\left\langle {\Delta
X_{i-1} \cdot \Delta X_i } \right\rangle }{\left\langle {\Delta
X_i^2 } \right\rangle }
\end{equation}

Where: $\Delta X_i \equiv X_i (\mbox{data})-X_i (\mbox{model})$

The scatter score will always be between 1 and -1. A scores close
to 1 occurs when all $\Delta X_i$ are approximately equal. In
practice, this represent a severe systematic error, where the
model is entirely above or entirely below the data. When there is
no systematic error, the data are distributed randomly around the
model, generating a scatter score approaching 0. Scores close to
-1, although theoretically possible when $\Delta X_i \simeq
-\Delta X_{i-1}$, can be considered unphysical in that they are
unlikely to occur through systematic or non-systematic errors.

\subsection{Waviness}

This is a special case of the scatter score (see previous subsection).
Here, we consider only data points in the light curve's plateau
(i.e. the region in the phased curve between the eclipsing dips,
where both stars are fully exposed). The Waviness score is the
scatter score of these data points around their median. The purpose
of this test is to get a model-independent measure of irregularities
in the binary brightness, out of eclipse. A large value may indicate
such effects as stellar elongation, spots, or flares.

\section{Density Estimation}

One of the most important criteria for selecting binaries for
follow-up is its stellar density. Unfortunately, the parameters
that can be extracted from the light curve fitting do not provide
us with enough information for deducing the density of any one of
the stars in the binary, but only a combined value. We define the
mean density as the sum of the stars' masses divided by the sum of
their volumes:

\begin{equation}
\bar {\rho }\equiv \frac{M_1+M_2}{\left( {4\pi /3} \right)\left(
{R_1^3+R_2^3 } \right)}=\frac{3\pi }{GP^2\left( {r_1^3 +r_2^3 }
\right)}\simeq \frac{0.01893\,g\,{cm}^{-3}}{P_{day}^2 \left(
{r_1^3+r_2^3 } \right)}\simeq \frac{0.01344\,
\rho_{\sun}}{P_{day}^2 \left( {r_1^3 + r_2^3} \right)}
\end{equation}

Where: $r_{1,2} \equiv R_{1,2} /a$ \ and from Kepler's law:
$a^3=G\left( {M_1+M_2} \right)\left( P/{2\pi} \right)^2$

It should be noted here that if the stars' have very different
sizes, their mean density will be dominated by the larger one,
according to the weighted average:

\begin{equation}
\bar{\rho } = \frac{\left( {r_1 / r_2} \right)^3\rho_1 + \rho_2
}{\left({r_1 / r_2} \right)^3 + 1}
\end{equation}

Similarly, assuming: $R_1 \ge R_2 $, the maximum possible density is:

\begin{equation}
\rho_{\max } \equiv \frac{M_1+M_2}{\left( {4\pi /3} \right)R_2^3 }
= \bar {\rho } \left( {1+\left( {r_1 /r_2} \right)^3}
\right)\simeq \frac{0.01893\,g\,{cm}^{-3}}{P_{day}^2 r_2^3 }\simeq
\frac{0.01344\,\rho_{\sun}}{P_{day}^2 r_2^3}
\end{equation}

Adding the assumption that the more dense star of the binary is
the less massive component, we can reduce the upper limit of its
density to $\rho_{\max } / 2$.

{}

\clearpage

\begin{deluxetable}{lcccc}
\tabletypesize{\scriptsize}
\tablecaption{FS Monocerotis (N = 249)}
\tablewidth{0pt}
\tablehead{\colhead{Parameters} & \colhead{Symbol} & \colhead{\citep{Lacy00}}
& \colhead{DEBiL best fit} & \colhead{Relative error}}
\startdata
Radius of primary (larger) star& $R_1/a$ & 0.2188 $\pm $ 0.0005& 0.222 $\pm $ 0.003& 1.3 {\%} \\
Radius of secondary (smaller) star& $R_2/a$ & 0.173 $\pm $ 0.003& 0.179 $\pm $ 0.006& 3.5 {\%} \\
Surface brightness ratio& $J_s$ & 0.903 $\pm $0.003& 0.916 $\pm $ 0.05& 1.5 {\%} \\
Orbital inclination& i& 87.48 $\pm $ 0.08& 87.86 $\pm $ 0.015& 0.4 {\%} \\
Eccentricity& e& 0.0 (fixed)& 0.001 $\pm $ 0.01& \\
\enddata
\label{tableLacyA}
\end{deluxetable}

\begin{deluxetable}{lcccc}
\tabletypesize{\scriptsize}
\tablecaption{WW Camelopardalis (N = 5759)}
\tablewidth{0pt}
\tablehead{\colhead{Parameters} & \colhead{Symbol} & \colhead{\citep{Lacy02}} &
\colhead{DEBiL best fit} & \colhead{Relative error}}
\startdata
Radius of primary (larger) star & $R_1/a$ & 0.168 $\pm $ 0.0013& 0.169 $\pm $ 0.018& 0.5 {\%} \\
Radius of secondary (smaller) star & $R_2/a$ & 0.159 $\pm $ 0.016& 0.165 $\pm $ 0.014& 3.4 {\%} \\
Surface brightness ratio & $J_s$ & 0.950 $\pm $ 0.003& 0.949 $\pm $ 0.08& 0.1 {\%} \\
Orbital inclination & i& 88.29 $\pm $ 0.06& 88.35 $\pm $ 0.03& 0.1 {\%} \\
Eccentricity& e& 0.0099 $\pm $ 0.0007& 0.01 $\pm $ 0.05& \\
\enddata
\label{tableLacyB}
\end{deluxetable}

\begin{deluxetable}{lcccc}
\tabletypesize{\scriptsize}
\tablecaption{BP Vulpeculae (N = 5236)}
\tablewidth{0pt}
\tablehead{\colhead{Parameters} & \colhead{Symbol} &
\colhead{\citep{Lacy03}} & \colhead{DEBiL best fit} &
\colhead{Relative error}}
\startdata
Radius of primary (larger) star& $R_1/a$ & 0.1899 $\pm $ 0.0008& 0.190 $\pm $ 0.006& 0.1 {\%} \\
Radius of secondary (smaller) star& $R_2/a$ & 0.161 $\pm $ 0.009& 0.166 $\pm $ 0.009& 3.1 {\%} \\
Surface brightness ratio& $J_s$ & 0.624 $\pm $ 0.0013& 0.614 $\pm $ 0.08& 1.5 {\%} \\
Orbital inclination& i& 86.71 $\pm $ 0.09& 86.50 $\pm $ 0.012& 0.2 {\%} \\
Eccentricity& e& 0.0355 $\pm $ 0.0005& 0.04 $\pm $ 0.03&\\
\enddata
\label{tableLacyC}
\end{deluxetable}

\begin{deluxetable}{cccccccccccccc}
\tabletypesize{\scriptsize}
\rotate
\tablecaption{Selected parameters from the DEBiL dataset of eclipsing binary systems in the galactic bulge.}
\tablewidth{0pt}
\tablehead{\colhead{Field} & \colhead{Object} & \colhead{Period} &
\colhead{e} & \colhead{$R_1/a$} & \colhead{$R_2/a$} &
\colhead{$I_1$ [mag.]} & \colhead{$I_2$ [mag.]} &
\colhead{$\sin(i)$} & \colhead{$t_0$ \tablenotemark{a}} &
\colhead{$\omega$ [deg.]} & \colhead{$\chi_\nu ^2$} &
\colhead{\begin{tabular}{c} Corrected \\ I [mag.] \tablenotemark{b} \end{tabular}} &
\colhead{\begin{tabular}{c} Corrected \\ V-I [mag.] \tablenotemark{b}\tablenotemark{c} \end{tabular}}}
\startdata
1& 39& 129.656& 0.074& 0.773& 0.118& 13.56& 17.21& 0.9208
& 0.665& 256.0& 1.13& 12.81& 1.93 \\
1& 45& 0.55677& 0.025& 0.589& 0.386& 17.36& 18.20& 0.9979
& 0.724& 271.4& 1.06& 16.21& -1000 \tablenotemark{d} \\
1& 53& 2.52158& 0.099& 0.412& 0.103& 12.20& 16.15& 1.0000
& 0.418& 91.4& 3.07& 11.38& 0.18 \\
1& 108& 1.53232& 0.005& 0.516& 0.301& 16.80& 19.12& 0.9143
& 0.497& 254.1& 0.93& 15.84& 0.48 \\
1& 112& 0.35658& 0.000& 0.514& 0.486& 17.90& 17.87& 0.9763
& 0.429& 40.6& 1.56& 16.36& 0.65 \\
1& 155& 0.96092& 0.014& 0.683& 0.303& 16.15& 17.84& 0.9203
& 0.199& 294.5& 1.12& 15.10& 0.45 \\
1& 183& 0.57793& 0.009& 0.555& 0.338& 17.99& 20.03& 0.9828
& 0.055& 56.4& 1.29& 17.14& 0.41 \\
1& 201& 0.67241& 0.101& 0.509& 0.246& 17.26& 20.08& 0.9971
& 0.473& 264.3& 0.87& 16.44& 0.32 \\
1& 202& 4.51345& 0.174& 0.309& 0.206& 17.58& 17.08& 0.9968
& 0.035& 270.9& 2.68& 15.82& 0.64 \\
1& 215& 0.48925& 0.000& 0.770& 0.230& 15.93& 18.62& 0.9212
& 0.870& 150.5& 1.54& 15.06& 0.20 \\
1& 221& 0.45013& 0.004& 0.558& 0.434& 18.23& 18.87& 0.9332
& 0.107& 100.7& 0.92& 16.92& 0.58 \\
1& 227& 122.879& 0.007& 0.700& 0.294& 16.69& 18.39& 1.0000
& 0.384& 224.3& 3.78& 15.77& 1.03 \\
1& 242& 0.28142& 0.003& 0.521& 0.476& 18.56& 18.67& 1.0000
& 0.274& 159.1& 1.38& 17.08& 0.77 \\
1& 246& 105.55& 0.011& 0.465& 0.071& 14.38& 18.25& 1.0000
& 0.280& 192.1& 3.68& 13.57& 1.19 \\
1& 257& 2.37881& 0.049& 0.244& 0.129& 17.25& 18.98& 0.9855
& 0.510& 87.5& 0.88& 16.33& 0.66 \\
\enddata
\tablecomments{This table is published in its entirety in the
electronic edition of the {\it Astrophysical Journal}. A portion
is shown here for guidance regarding its form and content.}
\tablenotetext{a}{Phased epoch of periastron: heliocentric Julian
date, minus 2450000.0, folded by the period.}
\tablenotetext{b}{Extinction corrected using the \citet{Sumi04}
extinction map of the galactic bulge.}
\tablenotetext{c}{The combined binary color was taken from \citep{Udalski02}.}
\tablenotetext{d}{The ``-1000'' values indicate missing magnitude or color data.}
\label{tableDEBiL}
\end{deluxetable}

\clearpage

\begin{deluxetable}{lccccccccccccc}
\tabletypesize{\scriptsize}
\rotate
\tablecaption{Extreme eclipsing binary system candidates}
\tablewidth{0pt}
\tablehead{\colhead{Category} & \colhead{Field} & \colhead{Object} &
\colhead{Period} & \colhead{e} & \colhead{$R_1/a$} & \colhead{$R_2/a$} &
\colhead{$I_1$ [mag.]} & \colhead{$I_2$ [mag.]} & \colhead{$\chi_\nu ^2$} &
\colhead{$\bar{\rho }$} & \colhead{$\rho_{\max }$} &
\colhead{\begin{tabular}{c} Corrected \\ I [mag.] \end{tabular}} &
\colhead{\begin{tabular}{c} Corrected \\ V-I [mag.] \end{tabular}}}
\startdata
High density&      21& 5952&   1.468& 0.001& 0.137& 0.039& 15.21& 17.78& 1.30&    3.372&   153.6& 14.26&  0.48 \\
High density&      23& 1774&   0.505& 0.018& 0.245& 0.162& 16.91& 17.41& 3.17&    3.925&   17.45& 14.87& -1000 \\
Low density&        3& 8264& 151.026& 0.023& 0.487& 0.228& 17.39& 18.95& 2.53& 0.000007& 0.00007& 15.46&  1.33 \\
Low density&       21& 2568& 186.496& 0.265& 0.472& 0.127& 15.28& 18.18& 1.82& 0.000005& 0.00026& 14.47&  1.04 \\
High eccentricity&  2& 547&    2.419& 0.231& 0.211& 0.077& 12.12& 14.66& 1.89&    0.330&   7.049& 11.47& -0.37 \\
High eccentricity& 38& 4059&   2.449& 0.454& 0.288& 0.179& 17.81& 20.65& 1.61&    0.107&   0.553& 16.94&  0.68 \\
Blue&              21& 3797&   2.653& 0.003& 0.193& 0.104& 12.33& 14.35& 3.03&    0.322&   2.373& 11.48& -0.37 \\
Blue&              30& 1778&   6.442& 0.084& 0.089& 0.028& 12.81& 15.53& 2.39&    0.627&   21.37& 11.81& -0.52 \\
Short period&      18& 3424&   0.179& 0.119& 0.670& 0.211& 16.73& 19.28& 2.10&    1.895&  62.905& 15.53&  0.13 \\
Short period\tablenotemark{e}&
                   46&  797&   0.198& 0.072& 0.444& 0.426& 16.67& 16.60& 2.80&    2.913&   6.208& 14.87&  1.36 \\
Short period&      49&  538&   0.228& 0.008& 0.728& 0.264& 16.09& 18.35& 1.07&    0.904&  19.929& 15.03&  0.31 \\
Short period&      42& 2087&   0.233& 0.021& 0.568& 0.336& 17.94& 19.40& 1.27&    1.580&   9.242& 16.68&  0.85 \\
\enddata

\tablenotetext{e}{After finding this candidate using our pipeline,
we identified it as the eclipsing binary BW3 V38, which was
discovered and extensively studied by \citet{Maceroni97}. For
consistency, we listed the DEBiL fitted parameters, though their
accuracy is considerably worse than what is currently available in
the literature \citep{Maceroni04}. Specifically, the DEBiL fits
for the binary components' radii are overestimated by $\sim$25\%
due to their tidal distortions (see \S\ref{subsecLimitations}).}

\label{tableExtreme}
\end{deluxetable}

\clearpage

\clearpage

\clearpage

\clearpage

\end{document}